\documentclass[11pt,english]{scrartcl}
\usepackage{lmodern}

\usepackage[T1]{fontenc}
\usepackage[a4paper]{geometry}
\usepackage{color}
\usepackage{rotfloat}
\usepackage{textcomp}
\usepackage{amsmath}
\usepackage{amssymb}
\usepackage{graphicx}
\usepackage{setspace}
\usepackage{esint}
\makeatletter

\providecommand{\tabularnewline}{\\}

\usepackage{amsthm}
\usepackage{psfrag}
\usepackage{epsfig}
\newtheorem{prop}{Proposition}
\newtheorem{defi}{Definition}
\usepackage[table]{xcolor}

\date{}

\newcommand{\diag}{\mathop{\mathrm{diag}}}

\makeatother

\usepackage{babel}
\begin{document}

\title{Kernel discriminant analysis and clustering\\
 with parsimonious Gaussian process models}

\author{C. Bouveyron$^{1}$, M. Fauvel$^{2}$ \& S. Girard$^{3}$}

\maketitle
\begin{center}
\vspace{-7ex}$^{1}$ Laboratoire SAMM, EA 4543, Universit\'e Paris
1 Panth\'eon-Sorbonne\\
 $^{2}$ DYNAFOR, UMR 1201, INRA \& Universit\'e de Toulouse\\
 $^{3}$ Equipe MISTIS, INRIA Rhône-Alpes \& LJK\\
FRANCE
\par\end{center}
\begin{abstract}
This work presents a family of parsimonious Gaussian process models
which allow to build, from a finite sample, a model-based classifier
in an infinite dimensional space. The proposed parsimonious models
are obtained by constraining the eigen-decomposition of the Gaussian
processes modeling each class. This allows in particular to use non-linear
mapping functions which project the observations into infinite dimensional
spaces. It is also demonstrated that the building of the classifier
can be directly done from the observation space through a kernel function.
The proposed classification method is thus able to classify data of
various types such as categorical data, functional data or networks.
Furthermore, it is possible to classify mixed data by combining different
kernels. The methodology is as well extended to the unsupervised classification
case. Experimental results on various data sets demonstrate the effectiveness
of the proposed method. 
\end{abstract}

\section{Introduction}

Classification is an important and useful statistical tool in all
scientific fields where decisions have to be made. Depending on the
availability of a learning data set, two main situations may happen:
supervised classification (also known as discriminant analysis) and
unsupervised classification (also known as clustering). Discriminant
analysis aims to build a classifier (or a decision rule) able to assign
an observation $x$ in an arbitrary space $E$ with unknown class
membership to one of $k$ known classes $C_{1},...,C_{k}$. For building
this supervised classifier, a learning dataset $\{(x_{1},z_{1}),...,(x_{n},z_{n})\}$
is used, where the observation $x_{\ell}\in E$ and $z_{\ell}\in\{1,...,k\}$
indicates the class belonging of the observation $x_{\ell}$. In a
slightly different context, clustering aims to directly partition
an incomplete dataset $\{x_{1},...,x_{n}\}$ into $k$ homogeneous
groups without any other information, \emph{i.e.,} assign to each
observation $x_{\ell}\in E$ its group membership $z_{\ell}\in\{1,...,k\}$.
Several intermediate situations exist, such as semi-supervised or
weakly-supervised classifications~\cite{ChaSchZie06}, but they will
not be considered here.

Since the pioneer work of Fisher~\cite{Fisher36}, a huge number
of supervised and unsupervised classification methods have been proposed
in order to deal with different types of data. Indeed, there exist
a wide variety of data such as quantitative, categorical and binary
data but also texts, functions, sequences, images and more recently
networks. As a practical example, biologists are frequently interested
in classifying biological sequences (DNA sequences, protein sequences),
natural language expressions (abstracts, gene mentioning), networks
(gene interactions, gene co-expression), images (cell imaging, tissue
classification) or structured data (gene structures, patient information).
The observation space $E$ can be therefore $\mathbb{R}^{p}$ if quantitative
data are considered, $L^{2}([0,1])$ if functional data are considered
(time series for example) or\textit{\textcolor{red}{{} }}$\mathcal{A}^{p}$,
where $\mathcal{A}$ is a finite alphabet, if the data at hand are
categorical (DNA sequences for example). Furthermore, the data to
classify can be a mixture of different data types: categorical and
quantitative data or categorical and network data for instance.

Classification methods can be split into two main families: generative
and discriminative techniques. On the one hand, generative techniques
model the data of each class with a probability distribution and deduce
the classification rule from this modeling. Model-based discriminant
analysis assumes that $\{x_{1},...,x_{n}\}$ are independent realizations
of a random vector $X$ on $E$ and that the class conditional distribution
of $X$ is parametric, \emph{i.e.} $f(x|z=i)=f_{i}(x;\theta_{i}).$
When $E={\mathbb{R}}^{p}$, among the possible parametric distributions
for $f_{i}$, the Gaussian distribution is often preferred and, in
this case, the marginal distribution of $X$ is therefore a mixture
of Gaussians: 
\[
f(x)=\sum_{i=1}^{k}\pi_{i}\phi(x;\mu_{i},\Sigma_{i}),
\]
where $\phi$ is the Gaussian density, $\pi_{i}$ is the prior probability
of the $i$th class, $\mu_{i}$ is the mean of the $i$th class and
$\Sigma_{i}$ is its covariance matrix. In such a case, the optimal
decision rule is called the \emph{maximum a posteriori} (MAP) rule
which assigns a new observation $x$ to the class which has the largest
posterior probability. Introducing the classification function $D_{i}(x)=\log|\Sigma_{i}|+(x-\mu_{i})^{t}\Sigma_{i}^{-1}(x-\mu_{i})-2\log(\pi_{i})$,
which can be rewritten as:
\begin{eqnarray}
D_{i}(x) & = & \sum_{j=1}^{p}\frac{1}{\lambda_{ij}}<x-\mu_{i},q_{ij}>_{\mathbb{R}^{p}}^{2}+\sum_{j=1}^{p}\log(\lambda_{ij})-2\log(\pi_{i}),\label{eq:classif-function}
\end{eqnarray}
where $q_{ij}$ and $\lambda_{ij}$ are respectively the $j$th eigenvector
and eigenvalue of $\Sigma_{i}$, it can be easily shown that the MAP
rule reduces to finding the label $i\in\{1,\dots,k\}$ for which $D_{i}(x)$
is the smallest. Estimation of model parameters is usually done by
maximum likelihood. This method is known as the quadratic discriminant
analysis (QDA), and, under the additional assumption that $\Sigma_{i}=\Sigma$
for all $i\in\{1,\dots,k\}$, it corresponds to the linear discriminant
analysis (LDA). A detailed overview on this topic can be found in~\cite{McLachlan92a}.
Recent extensions allowing to deal with high-dimensional data include~\cite{Bouveyron11,CSDA,BOUVEYRON:2007:INRIA-00176283:1,McLachlan2003,McNicholas2008,Montanari2010,Mur2010}.
Although model-based classification is usually enjoyed for its multiple
advantages, model-based discriminant analysis methods have however
two limiting characteristics. First, they are limited to quantitative
data and cannot process for instance qualitative or functional data.
Second, even in the case of quantitative data, the Gaussian assumption
may not be well-suited for the data at hand.

On the other hand, discriminative techniques directly build the classification
rule from the learning dataset. Among the discriminative classification
methods, kernel methods~\cite{1151.30007} are probably the most
popular and overcome some of the shortcomings of generative techniques.
Kernel methods are non-parametric algorithm and can be applied to
any data for which a kernel function can be defined. A kernel $K:E\times E\to\mathbb{R}$
is a positive definite function such as every evaluation can be written
as $K(x_{i},x_{j})=<\varphi(x_{i}),\varphi(x_{j})>_{\mathcal{H}}$,
with $x_{i},x_{j}\in E$, $\varphi$ a mapping function (called the
feature map), $\mathcal{H}$ a finite or infinite dimensional reproducing
kernel Hilbert space (the feature space) and $<\cdot,\cdot>_{\mathcal{H}}$
the dot product in $\mathcal{H}$. An advantage of using kernels is
the possibility of computing the dot product in the feature space
from the original input space without explicitly knowing $\varphi$
(kernel trick)~\cite{1151.30007}. Turning conventional learning
algorithms into kernel learning algorithms can be easily done if the
algorithms operate on the data only in terms of dot product. In particular,
the kernel trick is used to transform linear algorithms to non-linear
ones. Additionally, a nice property of kernel learning algorithms
is the possibility to deal with any kind of data. The only condition
is to be able to define a positive definite function over pairs of
elements to be classified~\cite{1151.30007}. For instance, kernel
functions can be defined on strings~\cite[Chap.~10  and 11]{kernel:method},
graphs~\cite{Smola09} or trees~\cite[Chap.~5]{kernel:biology}.
Many conventional linear algorithms have been turned to non-linear
algorithms thanks to kernels~\cite{Scholkopf:2001:LKS:559923}. For
instance, a kernelized version of principal component analysis (KPCA)
has been proposed in~\cite{doi:10.1162/089976698300017467}. Mika
\emph{et al.} have also proposed kernel Fisher discriminant (KFD)
as a non-linear version of FDA which only relies on kernel evaluations~\cite{788121}.
A kernelized Gaussian mixture model (KGMM) has been proposed in~\cite{Dundar04towardan}
for the supervised classification of hyperspectral data. But, due
to computational considerations, the authors have introduced a strong
assumption: the classes share the same covariance matrix in the feature
space. However, the method still needs to be regularized. Recently,
pseudo-inverse and ridge regularization have been proposed to define
a kernel quadratic classifier where classes have their own covariance
matrices~\cite{4711053}. In all these cases, a benefit is found
by using the kernel version rather than the original algorithm. KPCA
shows better results results than PCA in terms of reconstruction errors
for image denoising~\cite{1471703}. Kernel GMM provides better accuracy
than conventional GMM for the classification of hyperspectral images~\cite{Dundar04towardan}.
Let us however highlight that the kernel version involves the inversion
of a kernel matrix, \emph{i.e.}, a $n\times n$ matrix estimated with
only $n$ samples. Usually, the kernel matrix is ill-conditioned and
regularization is needed, while sometimes a simplified model is required
too. Thus, it may limit the effectiveness of the kernel version. In
addition, and conversely to model-based techniques, the classification
results provided by kernel methods are unfortunately difficult to
interpret which would be useful in many application domains.

In this work, we propose to adapt model-based methods for the classification
of any kind of data by working in a feature space of high or even
infinite dimensional space. To this end, we propose a family of parsimonious
Gaussian process models which allow to build, from a finite sample,
a model-based classifier in a infinite dimensional space. It will
be demonstrated that the building of the classifier can be directly
done from the observation space through the so-called ``kernel trick''.
The proposed classification method will be thus able to classify data
of various types (categorical data, mixed data, functional data, networks,
...). The methodology is as well extended to the unsupervised classification
case (clustering).

The paper is organized as follows. Section~\ref{sec:model:khdda}
presents the context of our study and introduces the family of parsimonious
Gaussian process models. The inference aspects are addressed in Section~\ref{sec:infer:khdda}.
It is also demonstrated in this section that the proposed method can
work directly from the observation space through a kernel. Section~\ref{sec:clustering}
is dedicated to some special cases and to the extension to the unsupervised
framework. Experimental comparisons with state-of-the-art kernel methods
are presented in Section~\ref{sec:simul} as well as applications
of the proposed methodologies to various types of data including functional,
categorical, mixed and network data. Some concluding remarks are given
in Section~\ref{sec:conclu} and proofs are postponed to the appendix.

\section{Classification with parsimonious Gaussian process models\label{sec:model:khdda}}

In this section, we first define the context of our approach and exhibit
the associated computational problems. Then, a parsimonious parameterization
of Gaussian processes is proposed in order to overcome the highlighted
computational issues.

\subsection{Classification with Gaussian processes}

Let us consider a learning set $\{(x_{1},z_{1}),...,(x_{n},z_{n})\}$
where $\{x_{1},...,x_{n}\}\subset E$ are assumed to be independent
realizations of a, possibly non-quantitative and non-Gaussian, random
variable $X$. The class labels $\{z_{1},...,z_{n}\}$ are assumed
to be realizations of a discrete random variable $Z\in\{1,...,k\}$.
It indicates the memberships of the learning data to the $k$ classes
denoted by $C_{1},\dots,C_{k}$, \emph{i.e.,} $z_{\ell}=i$ indicates
that $x_{\ell}$ belongs to $C_{i}$.

Let us assume that there exists a non-linear mapping $\varphi$ such
that $Y=\varphi(X)$ is, conditionally on $Z=i$, a Gaussian process
on $[0,1]$ with mean $\mu_{i}$ and continuous covariance function
$\Sigma_{i}$. More specifically, one has $\mu_{i}(t)={\mathbb{E}}(Y(t)|Z=i)$
and $\Sigma_{i}(s,t)={\mathbb{E}}(Y(s)Y(t)|Z=i)-\mu_{i}(t)\mu_{i}(s)$.
It is then well-known~\cite{Shorack86} that, for all $i=1,\dots,k$,
there exist positive eigenvalues (sorted in decreasing order) $\{\lambda_{ij}\}_{j\geq1}$,
together with eigenvector functions $\{q_{ij}(.)\}_{j\geq1}$ continuous
on $[0,1]$, such that 
\[
\Sigma_{i}(s,t)=\sum_{j=1}^{\infty}\lambda_{ij}q_{ij}(s)q_{ij}(t),
\]
 where the series is uniformly convergent on $[0,1]^{2}$. Moreover,
the eigenvector functions are orthonomal in $L^{2}([0,1])$ for the
dot product $<f,g>_{L_{2}}=\int_{0}^{1}f(t)g(t)dt$. It is then easily
seen, that, for all $r\geq1$ and $i\in\{1,\dots,k\}$, the random
vector on ${\mathbb{R}}^{r}$ defined by $\{<Y,q_{ij}>_{L_{2}}\}_{j=1,\dots,r}$
is, conditionally on $Z=i$, Gaussian with mean $\{<\mu_{i},q_{ij}>\}_{j=1,\dots,r}$
and covariance matrix $\diag(\lambda_{i1},\dots,\lambda_{ir})$. To
classify a new observation $x$, we therefore propose to apply the
Gaussian classification function~(\ref{eq:classif-function}) to
$\varphi(x)$: 
\[
D_{i}(\varphi(x))=\sum_{j=1}^{r}\frac{1}{\lambda_{ij}}<\varphi(x)-\mu_{i},q_{ij}>_{L_{2}}^{2}+\sum_{j=1}^{r}\log(\lambda_{ij})-2\log(\pi_{i}).
\]
From a theoretical point of view, if the Gaussian process is non degenerated,
one should use $r=+\infty$. In practice, $r$ has to be large in
order not to loose to much information on the Gaussian process. Unfortunately,
in this case, the above quantities cannot be estimated from a finite
sample set. Indeed, only a part of the classification function can
be actually computed from a finite sample set: 
\begin{eqnarray*}
D_{i}(\varphi(x)) & = & \underbrace{\sum_{j=1}^{r_{i}}\frac{1}{\lambda_{ij}}<\varphi(x)-\mu_{i},q_{ij}>_{L_{2}}^{2}+\sum_{j=1}^{r_{i}}\log(\lambda_{ij})-2\log(\pi_{i})}_{\text{computable quantity}}\\
 & + & \underbrace{\sum_{j=r_{i}+1}^{r}\frac{1}{\lambda_{ij}}<\varphi(x)-\mu_{i},q_{ij}>_{L_{2}}^{2}+\sum_{j=r_{i}+1}^{r}\log(\lambda_{ij})}_{\text{non computable quantity}},
\end{eqnarray*}
where $r_{i}=\min(n_{i},r)$ and $n_{i}=\mathrm{Card}\left(C_{i}\right)$.
Consequently, the Gaussian model cannot be used directly in the feature
space to classify data if $r>n_{i}$ for $i=1,...,k$.

\subsection{A parsimonious Gaussian process model}

To overcome the computation problem highlighted above, it is proposed
here to use in the feature space a parsimonious model for the Gaussian
process modeling each class\textit{\textcolor{red}{.}} Following the
idea of~\cite{BOUVEYRON:2007:INRIA-00176283:1}, we constrain the
eigen-decomposition of the Gaussian processes as follows.

\begin{defi}A parsimonious Gaussian process model is a Gaussian process
$Y$ for which, conditionally to $Z=i$, the eigen-decomposition of
its covariance operator $\Sigma_{i}$ is such that:\begin{enumerate}

\item[{(A1)}] it exists a dimension $r<+\infty$ such that $\lambda_{ij}=0$
for $j\geq r$ and for all $i=1,...,k,$

\item[{(A2)}] it exists a dimension $d_{i}<\min\{r,n_{i}\}$ such
that $\lambda_{ij}=\lambda$ for $d_{i}<j<r$ and for all $i=1,...,k$.\end{enumerate}

\end{defi}

It is worth noticing that $r$ can be as large as it is desired, as
long it is finite, and in particular $r$ can be much larger than
$n_{i}$, for any $i=1,...,k$. From a practical point of view, this
modeling can be viewed as assuming that the data of each class live
in a specific subspace of the feature space. The variance of the actual
data of the $i$th group is modeled by the parameters $\lambda_{i1},...,\lambda_{id_{i}}$
and the variance of the noise is modeled by $\lambda$. This assumption
amounts to supposing that the noise is homoscedastic and its variance
is common to all the classes. The dimension $d_{i}$ can be considered
as well as the intrinsic dimension of the latent subspace of the $i$th
group in the feature space. This model is referred to by pgp${\cal M}_{0}$
(or ${\cal M}_{0}$ for short) hereafter. With these assumptions,
we have the following result.

\begin{prop}\label{propun}Letting $d_{\max}=\max(d_{1},...,d_{k})$,
the classification function $D_{i}$ can be written as follows in
the case of a parsimonious Gaussian process model pgp$\mathcal{M}$:
\begin{eqnarray}
D_{i}(\varphi(x)) & = & \sum_{j=1}^{d_{i}}\left(\frac{1}{\lambda_{ij}}-\frac{1}{\lambda}\right)<\varphi(x)-\mu_{i},q_{ij}>_{L_{2}}^{2}+\frac{1}{\lambda}||\varphi(x)-\mu_{i}||_{L_{2}}^{2}\nonumber \\
 & + & \sum_{j=1}^{d_{i}}\log(\lambda_{ij})+(d_{\max}-d_{i})\log(\lambda)-2\log(\pi_{i})+\gamma,\label{eq:classif-function-1}
\end{eqnarray}
 where $\gamma$ is a constant term which does not depend on the index
$i$ of the class.\end{prop}

At this point, it is important to notice that the classification function
$D_{i}$ depends only on the eigenvectors associated with the $d_{i}$
largest eigenvalues of $\Sigma_{i}$. This estimation is now possible
due to the inequality $d_{i}<n_{i}$ for $i=1,...,k$. Furthermore,
the computation of the classification function does not depend any
more on the parameter $r$. As shown in the next section, it is possible
to reformulate the classification function such that it does not depend
either on the mapping function $\varphi$.

\subsection{Submodels of the parsimonious model}

By fixing some parameters to be common within or between classes,
it is possible to obtain particular models which correspond to different
regularizations. Table~\ref{Table_models} presents the 8 additional
models which can be obtained by constraining the parameters of model
${\cal M}_{0}$. For instance, fixing the dimensions $d_{i}$ to be
common between the classes yields the model~${\cal M}_{1}$. Similarly,
fixing the first $d_{i}$ eigenvalues to be common within each class,
we obtain the more restricted model ${\cal M}_{2}$. It is also possible
to constrain the first $d_{i}$ eigenvalues to be common between the
classes (models ${\cal M}_{4}$ and ${\cal M}_{7}$), and within and
between the classes (models ${\cal M}_{5}$, ${\cal M}_{6}$ and ${\cal M}_{8}$).
This family of 9 parsimonious models should allow the proposed classification
method to fit into various situations.

\noindent 
\begin{table}
\begin{centering}
{\small%
\begin{tabular}{|c|c|c|c|c|}
\hline 
Model  & %
\begin{tabular}{c}
Variance inside\tabularnewline
the subspace $F_{i}$ \tabularnewline
\end{tabular} & %
\begin{tabular}{c}
Variance\tabularnewline
outside $F_{i}$ \tabularnewline
\end{tabular} & %
\begin{tabular}{c}
Subspace \tabularnewline
orientation $Q_{i}$ \tabularnewline
\end{tabular} & %
\begin{tabular}{c}
Intrinsic\tabularnewline
dimension $d_{i}$\tabularnewline
\end{tabular}\tabularnewline
\hline 
\hline 
${\cal M}_{0}$  & Free  & Common  & Free  & Free\tabularnewline
\hline 
${\cal M}_{1}$  & Free  & Common  & Free  & Common\tabularnewline
\hline 
${\cal M}_{2}$  & Common within groups  & Common  & Free  & Free\tabularnewline
\hline 
${\cal M}_{3}$  & Common within groups  & Common  & Free  & Common\tabularnewline
\hline 
${\cal M}_{4}$  & Common between groups  & Common  & Free  & Common\tabularnewline
\hline 
${\cal M}_{5}$  & Common within and between groups  & Common  & Free  & Free\tabularnewline
\hline 
${\cal M}_{6}$  & Common within and between groups  & Common  & Free  & Common\tabularnewline
\hline 
${\cal M}_{7}$  & Common between groups  & Common  & Common  & Common\tabularnewline
\hline 
${\cal M}_{8}$  & Common within and between groups  & Common  & Common  & Common\tabularnewline
\hline 
\end{tabular} }
\par\end{centering}

\caption{\label{Table_models}List of the submodels of the parsimonious Gaussian
process model (referred to by ${\cal M}_{0}$ here).}
\end{table}

\section{Model inference and classification with a kernel\label{sec:infer:khdda}}

This section focuses on the inference of the parsimonious models proposed
above and on the classification of new observations through a kernel.
Model inference is only presented for the model $\mathcal{M}_{0}$
since inference for the other parsimonious models is similar. Estimation
of intrinsic dimensions and visualization in the feature subspaces
are also discussed.

\subsection{Estimation of model parameters\label{sec:theo}}

In the model-based classification context, parameters are usually
estimated by their empirical counterparts~\cite{McLachlan92a} which
conduces, in the present case, to estimate the proportions $\pi_{i}$
by $\hat{\pi}_{i}={n_{i}}/{n}$ and the mean function $\mu_{i}$ by
$\hat{\mu}_{i}(t)={\displaystyle \frac{1}{n_{i}}\sum_{x_{j}\in C_{i}}\varphi(x_{j})(t)}$.
Regarding the covariance operator, the eigenvalue $\lambda_{ij}$
and the eigenvector $q_{ij}$ are respectively estimated by the $j$th
largest eigenvalue $\hat{\lambda}_{ij}$ and its associated eigenvector
function $\hat{q}_{ij}$ of the empirical covariance operator $\hat{\Sigma}_{i}$:
\[
\hat{\Sigma}_{i}(s,t)=\frac{1}{n_{i}}\sum_{x_{\ell}\in C_{i}}\varphi(x_{\ell})(s)\varphi(x_{\ell})(t)-\hat{\mu}_{i}(s)\hat{\mu}_{i}(t).
\]
Finally, the estimator of $\lambda$ is: 
\begin{equation}
\hat{\lambda}=\frac{1}{\sum_{i=1}^{k}\hat{\pi}_{i}\left(r-d_{i}\right)}\sum_{i=1}^{k}\hat{\pi}_{i}\left(\mathrm{trace}(\hat{\Sigma}_{i})-\sum_{j=1}^{d_{i}}\hat{\lambda}_{ij}\right).\label{eq:lambda}
\end{equation}
Using the plug-in method, the estimated classification function $\hat{D}_{i}$
can be written as follows:
\begin{eqnarray}
\hat{D}_{i}(\varphi(x)) & = & {\displaystyle \sum_{j=1}^{d_{i}}\left(\frac{1}{\hat{\lambda}_{ij}}-\frac{1}{\hat{\lambda}}\right)<\varphi(x)-\hat{\mu}_{i},\hat{q}_{ij}>_{L_{2}}^{2}+\frac{1}{\hat{\lambda}}||\varphi(x)-\hat{\mu}_{i}||_{L_{2}}^{2}}\nonumber \\
 & + & {\displaystyle \sum_{j=1}^{d_{i}}\log(\hat{\lambda}_{ij})+(d_{\max}-d_{i})\log(\hat{\lambda})-2\log(\hat{\pi}_{i}).\label{eq:estim-Di}}
\end{eqnarray}
However, as we can see, the estimated classification function $\hat{D}_{i}$
still depends on the function $\varphi$ and therefore requires computations
in the feature space. However, since all these computations involve
dot products, it will be shown in the next paragraph that the estimated
classification function can be computed without explicit knowledge
of $\varphi$ through a kernel function.

\subsection{Estimation of the classification function through a kernel}

Kernel methods are all based on the so-called ``kernel trick'' which
allows the computation of the classifier in the observation space
through a kernel $K$. Let us therefore introduce the kernel $K:E\times E\to{\mathbb{R}}$
defined as $K(x,y)=<\varphi(x),\varphi(y)>_{L_{2}}$ and $\rho_{i}:E\times E\to{\mathbb{R}}$
defined as $\rho_{i}(x,y)=<\varphi(x)-\mu_{i},\varphi(y)-\mu_{i}>_{L_{2}}$.
In the following, it is shown that the classification function $D_{i}$
only involves $\rho_{i}$ which can be computed using $K$: 
\begin{align}
\rho_{i}(x,y) & =\frac{1}{n_{i}^{2}}\sum_{x_{\ell},x_{\ell'}\in C_{i}}<\varphi(x)-\varphi(x_{\ell}),\varphi(y)-\varphi(x_{\ell'})>_{L_{2}}\\
 & =K(x,y)-\frac{1}{n_{i}}\sum_{x_{\ell}\in C_{i}}(K(x_{\ell},y)+K(x,x_{\ell}))+\frac{1}{n_{i}^{2}}\sum_{x_{\ell},x_{\ell'}\in C_{i}}K(x_{\ell},x_{\ell'}).
\end{align}
 For each class $C_{i}$, let us introduce the $n_{i}\times n_{i}$
symmetric matrix $M_{i}$ defined by: 
\[
(M_{i})_{\ell,\ell'}=\frac{\rho_{i}(x_{\ell},x_{\ell'})}{n_{i}}.
\]
With these notations, we have the following result.

\begin{prop} \label{propdeux} For $i=1,\dots,k$, the estimated
classification function can be computed, in the case of the model
$\mathcal{M}_{0}$, as follows: 
\begin{align*}
\hat{D}_{i}(\varphi(x)) & =\frac{1}{n_{i}}\sum_{j=1}^{d_{i}}\frac{1}{\hat{\lambda}_{ij}}\left(\frac{1}{\hat{\lambda}_{ij}}-\frac{1}{\hat{\lambda}}\right)\left(\sum_{x_{\ell}\in C_{i}}\beta_{ij\ell}\rho_{i}(x,x_{\ell})\right)^{2}+\frac{1}{\hat{\lambda}}\rho_{i}(x,x)\\
 & +\sum_{j=1}^{d_{i}}\log(\hat{\lambda}_{ij})+(d_{\max}-d_{i})\log(\hat{\lambda})-2\log(\hat{\pi}_{i}),
\end{align*}
where, for $j=1,\dots,d_{i}$, $\beta_{ij}$ is the normed eigenvector
associated to the $j$th largest eigenvalue $\hat{\lambda}_{ij}$
of $M_{i}$ and $\hat{\lambda}=1/\sum_{i=1}^{k}\hat{\pi_{i}}(r_{i}-d_{i})\times\sum_{i=1}^{k}\hat{\pi}_{i}\left(\mathrm{trace}(M_{i})-\sum_{j=1}^{d_{i}}\hat{\lambda}_{ij}\right).$
\end{prop}

It thus appears that each new sample point $x$ can be assigned to
the class $C_{i}$ with the smallest value of the classification function
without knowledge of $\varphi$. The methodology based on Proposition~\ref{propdeux}
is referred to pgpDA in the sequel. In practice, the value of $r_{i}$
depends on the chosen kernel (see Table~\ref{tab:ri} for examples).

\noindent 
\begin{table}
\centering%
\begin{tabular}{|l|c|c|}
\hline 
Kernels  & $K(x,y)$  & $r_{i}$\tabularnewline
\hline 
Linear & $<x,y>_{L_{2}}$  & $\min(n_{i},p)$\tabularnewline
Gaussian & $\exp\left(-\frac{\|x-y\|_{L_{2}}^{2}}{2\sigma^{2}}\right)$  & $n_{i}$ \tabularnewline
Polynomial & $\left(<x,y>_{L_{2}}+1\right)^{q}$  & $\min\left(n_{i},\binom{p+q}{p}\right)$\tabularnewline
\hline 
\end{tabular}\caption{\label{tab:ri}Dimension $r_{i}$ for several kernels.}
\end{table}

\subsection{Intrinsic dimension estimation }

The estimation of the intrinsic dimension of a dataset is a difficult
problem which occurs frequently in data analysis, such as in principal
component analysis. A classical solution in PCA is to look for a break
in the eigenvalue scree of the covariance matrix. This strategy relies
on the fact that the $j$th eigenvalue of the covariance matrix corresponds
to the fraction of the full variance carried by the $j$th eigenvector
of this matrix. Since, in our case, the class conditional matrix $M_{i}$
shares with the empirical covariance operator of the associated class
its largest eigenvalues, we propose to use a similar strategy based
on the eigenvalue scree of the matrices $M_{i}$ to estimate $d_{i}$,
$i=1,...,k$. More precisely, we propose to make use of the scree-test
of Cattell~\cite{Cattell66} for estimating the class specific dimension
$d_{i}$, $i=1,...,k$. For each class, the selected dimension is
the one for which the subsequent eigenvalues differences are smaller
than a threshold which can be tuned by cross-validation for instance.

\subsection{Visualization in the feature subspaces}

An interesting advantage of the approach is to allow the visualization
of the data in subspaces of the feature space. Indeed, even though
the chosen mapping function is associated with a space of very high
or infinite dimension, the proposed methodology models and classifies
the data in low-dimensional subspaces of the feature space. It is
therefore possible to visualize the projection of the mapped data
on the feature subspaces of each class using Equation~(\ref{projected-data})
of the appendix. The projection of $\varphi(x)$ on the $j$th axis
of the class $C_{i}$ is therefore given by: 
\[
P_{ij}(\varphi(x)):=<\varphi(x)-\hat{\mu}_{i},\hat{q}_{ij}>=\frac{1}{\sqrt{n_{i}\hat{\lambda}_{ij}}}\sum_{x_{\ell}\in C_{i}}\beta_{ij\ell}\rho_{i}(x,x_{\ell}).
\]
 Thus, even if the observations are non quantitative, it is possible
to visualize their projections in the feature subspaces of the classes
which are quantitative spaces.

\section{Particular cases and extension to clustering\label{sec:clustering}}

The methodology proposed in the previous section is made very general
by the large choice for the mapping function $\varphi(x)$. We focus
in this section on two specific choices for $\varphi(x)$ for which
the direct calculation of the classification rule is possible. An
extension to unsupervised classification is also considered through
the use of an EM algorithm.

\subsection{Case of the linear kernel for quantitative data}

In the case of quantitative data, $E={\mathbb{R}}^{p}$ and one can
choose $\varphi(x)=x$ associated to the standard scalar product which
gives rise to the linear kernel $K(x,y)=x^{t}y$. In such a framework,
the estimated classification function can be simplified as follows:
\begin{prop} \label{proptrois} If $E={\mathbb{R}}^{p}$ and $K(x,y)=x^{t}y$
then, for $i=1,\dots,k$, the estimated classification function reduces
to 
\begin{align*}
\hat{D}_{i}(x) & =\sum_{j=1}^{d_{i}}\left(\frac{1}{\hat{\lambda}_{ij}}-\frac{1}{\hat{\lambda}}\right)\left(\hat{q}_{ij}^{t}(x-\hat{\mu}_{i})\right)^{2}+\frac{1}{\hat{\lambda}}||x-\hat{\mu}_{i}||_{\mathbb{R}^{p}}^{2}\\
 & +\sum_{j=1}^{d_{i}}\log(\hat{\lambda}_{ij})+(d_{\max}-d_{i})\log(\hat{\lambda})-2\log(\hat{\pi}_{i}).
\end{align*}
 where $\hat{\mu}_{i}$ is the empirical mean of the class $C_{i}$,
$\hat{q}_{ij}$ is the eigenvector of the empirical covariance matrix
$\hat{\Sigma}_{i}$ associated to the $j$th largest eigenvalue $\hat{\lambda}_{ij}$
and $\hat{\lambda}$ is given by~(\ref{eq:lambda}).\end{prop}

It appears that the estimated classification function reduces to the
one of the HDDA method~\cite{BOUVEYRON:2007:INRIA-00176283:1} with
the model $[a_{ij}bQ_{i}d]$ which has constraints similar to ${\cal M}_{0}$.
Therefore, the methodology proposed in this work partially encompasses
the method HDDA.

\subsection{Case of functional data\label{sub:Case-of-functional}}

\noindent Let us consider now functional data observed in $E=L^{2}([0,1])$.
Let $(b_{j})_{j\geq1}$ be a basis of $L^{2}([0,1])$ and $F={\mathbb{R}}^{L}$
where $L$ is a given integer. For all $\ell=1,\dots,L$, the projection
of a function $x$ on the $j$th basis function is computed as 
\[
\gamma_{j}(x)=\int_{0}^{1}x(t)b_{j}(t)dt
\]
 and $\gamma(x):=(\gamma_{j}(x))_{j=1,\dots,L}$. Let $B$ the $L\times L$
Gram matrix associated to the basis: 
\[
B_{j\ell}=\int_{0}^{1}b_{j}(t)b_{\ell}(t)dt,
\]
 and consider the associated scalar product defined by $<u,v>=u^{t}Bv$
for all $u,v\in{\mathbb{R}}^{L}$. One can then choose $\varphi(x)=B^{-1}\gamma(x)$
and $K(x,y)=\gamma(x)^{t}B^{-1}\gamma(y)$ leading to a simple estimated
classification function.

\noindent \begin{prop} \label{propquatre} Let $E=L^{2}([0,1])$
and $K(x,y)=\gamma(x)^{t}B^{-1}\gamma(y)$. Introduce, for $i=1,\dots,k$,
the $L\times L$ covariance matrix of the $\gamma(x_{j})$ when $x_{j}\in C_{i}$:
\[
\hat{\Sigma}_{i}=\frac{1}{n_{i}}\sum_{x_{\ell}\in C_{i}}(\gamma(x_{\ell})-\bar{\gamma}_{i})(\gamma(x_{\ell})-\bar{\gamma}_{i})^{t}\mbox{ where }\bar{\gamma}_{i}=\frac{1}{n_{i}}\sum_{x_{j}\in C_{i}}\gamma(x_{j})
\]
 Then, for $i=1,\dots,k$, the estimated classification function reduces
to 
\begin{align*}
\hat{D}_{i}(\varphi(x)) & =\sum_{j=1}^{d_{i}}\left(\frac{1}{\hat{\lambda}_{ij}}-\frac{1}{\hat{\lambda}}\right)\left(\hat{q}_{ij}^{t}(\gamma(x)-\bar{\gamma}_{i})\right)^{2}+\frac{1}{\hat{\lambda}}(\gamma(x)-\bar{\gamma}_{i})^{t}B^{-1}(\gamma(x)-\bar{\gamma}_{i})\\
 & +\sum_{j=1}^{d_{i}}\log(\hat{\lambda}_{ij})+(d_{\max}-d_{i})\log(\hat{\lambda})-2\log(\hat{\pi}_{i}),
\end{align*}
 where $\hat{q}_{ij}$ and $\hat{\lambda}_{ij}$ are respectively
the $j$th normed eigenvector and eigenvalue of the matrix $B^{-1}\hat{\Sigma}_{i}$
and $\hat{\lambda}$ is given by~(\ref{eq:lambda}).\end{prop}

\noindent Remark that $B^{-1}\hat{\Sigma}_{i}$ coincides with the
matrix of interest in functional PCA~\cite[Chap.~8.4]{Ram2005} and
that, if the basis is orthogonal, then $B$ is the identity matrix.
Notice that the proposed method therefore encompasses as well the
model proposed in~\cite{Bouveyron11b} for the clustering of functional
data.

\subsection{Extension to unsupervised classification}

\noindent Since the previous section has demonstrated the possibility
to use the Gaussian classification function in the feature space,
it is also possible to extend its use to unsupervised classification
(also known as clustering). Indeed, in the model-based classification
context, the unsupervised and supervised cases mainly differ in the
manner to estimate the parameters of the model. The clustering task
aims to form $k$ homogeneous groups from a set of $n$ observations
$\{x_{1},...,x_{n}\}$ without any prior information about their group
memberships. Since the labels are not available, it is not possible
in this case to directly estimate the model parameters. In such a
context, the expectation-maximization (EM) algorithm~\cite{Dempster77}
is frequently used. As a consequence, the use of the EM algorithm
allows to both estimate the model parameters and predict the class
memberships of the observations at hand. In the case of the parsimonious
model~$\mathcal{M}_{0}$ introduced above, the EM algorithm takes
the following form:

\paragraph{The E step}

This first step reduces, at iteration $q$, to the computation of
$t_{ij}^{(q)}={\mathbb{E}}(Z_{j}=i|x_{j},\theta^{(q-1)})$, for $j=1,\dots,n$
and $i=1,\dots,k$, conditionally on the current value of the model
parameter $\theta^{(q-1)}$: 
\begin{eqnarray}
t_{ij}^{(q)}=1/\sum_{\ell=1}^{k}\exp\left(D_{i}^{(q-1)}(\varphi(x_{j}))-D_{\ell}^{(q-1)}(\varphi(x_{j}))\right),
\end{eqnarray}
 where 
\begin{align*}
D_{i}^{(q-1)}(\varphi(x)) & =\frac{1}{n_{i}}\sum_{j=1}^{d_{i}}\frac{1}{\hat{\lambda}_{ij}^{(q-1)}}\left(\frac{1}{\hat{\lambda}_{ij}^{(q-1)}}-\frac{1}{\hat{\lambda}^{(q-1)}}\right)\left(\sum_{\ell=1}^{n}\beta_{ij\ell}\sqrt{t_{i\ell}}\rho_{i}^{(q-1)}(x,x_{\ell})\right)^{2}\\
 & +\frac{1}{\hat{\lambda}^{(q-1)}}\rho_{i}^{(q-1)}(x,x)+\sum_{j=1}^{d_{i}}\log(\hat{\lambda}_{ij}^{(q-1)})+(d_{\max}-d_{i})\log(\hat{\lambda}^{(q-1)})-2\log(\hat{\pi}_{i}^{(q-1)}).
\end{align*}
is the Gaussian classification function associated with the model
parameters estimated in the M step at iteration $q-1$. This result
can be proved by substituting Equation~(\ref{eqvect}) in the proof
of Proposition~\ref{propdeux} by: 
\begin{equation}
\hat{q}_{ij}=\frac{1}{\sqrt{n_{i}\hat{\lambda}_{ij}}}\sum_{x_{\ell}\in C_{i}}\beta_{ij\ell}\sqrt{t_{\ell i}}(\varphi(x_{\ell})-\hat{\mu}_{i}).
\end{equation}

\paragraph{The M step}

This second step estimates the model parameters conditionally on the
posterior probabilities $t_{ij}^{(q)}$ computed in the previous step.
In practice, this step reduces to update the estimate of model parameters
according to the following formula: 
\begin{itemize}
\item mixture proportions are estimated by $\hat{\pi}_{i}^{(q)}=n_{i}^{(q)}/n$
where $n_{i}^{(q)}=\sum_{j=1}^{n}t_{ij}^{(q)}$, 
\item parameters $\lambda_{ij}$, $\lambda$, $\beta_{ij}$ and $d_{i}$
are estimated at iteration $q$ using the formula given in Proposition~\ref{propdeux}
but where the matrix $M_{i}$ is now a $n\times n$ matrix, recomputed
at each iteration $q$, and such that, for $i=1,...,k$ and $\ell,\ell'=1,...,n$:
\[
\left(M_{i}^{(q)}\right)_{\ell,\ell'}=\frac{\sqrt{t_{i\ell}^{(q)}t_{i\ell'}^{(q)}}}{n_{i}^{(q)}}\rho_{i}^{(q)}(x_{\ell},x_{\ell'})
\]
 where $\rho_{i}^{(q)}(x_{\ell},x_{\ell'})$ can be computed through
the kernel $K$ as follows: 
\begin{align*}
\rho_{i}^{(q)}(x_{\ell},x_{\ell'})= & K(x_{\ell},x_{\ell'})-\frac{1}{n_{i}^{(q)}}\sum_{j=1}^{n}t_{ji}^{(q)}\left(K(x_{j},x_{\ell})+K(x_{\ell'},x_{j})\right)\\
 & +\frac{1}{(n_{i}^{(q)})^{2}}\sum_{j,j'=1}^{n}t_{ji}^{(q)}t_{j'i}^{(q)}K(x_{j},x_{j'}).
\end{align*}
 
\end{itemize}
The clustering algorithm associated with this methodology will be
denoted to by pgpEM in the following.

\section{Benchmark study and applications to non-quantitative data\label{sec:simul}}

In this section, numeral experiments and comparisons are conducted
on real-world data sets to highlight the main features of the pgpDA
and pgpEM methods.

\subsection{Benchmark study on quantitative data}

We focus here on the comparison of pgpDA with state-of-the-art methods.
To that end, two kernel generative classifiers are considered, kernel
Fisher discriminant analysis (KFD)~\cite{788121} and kernel QDA
(KQDA)~\cite{Dundar04towardan}, and one kernel discriminative classifier,
support vector machine (SVM)~\cite{Scholkopf:2001:LKS:559923}. The
Gaussian kernel is used once again in the experiments for all methods,
including pgpDA. Since real-world problems are considered, all the
hyper-parameters of the classifiers have been tuned using 5-fold cross-validation.

Six data sets from the UCI Machine Learning Repository (\emph{http://archive.ics.uci.edu/ml/})
have been selected: glass, ionosphere, iris, sonar, USPS and wine.
We selected these data sets because they represent a wide range of
situations in term of number of observations $n$, number of variables
$p$ and number of groups $k$. The USPS dataset has been modified
to focus on discriminating the three most difficult classes to classify,
namely the classes of the digits 3, 5 and 8. This dataset has been
called USPS~358. The main feature of the data sets are described
in Table~\ref{tab:data}.

\begin{table}
\begin{centering}
\begin{tabular}{|c|c|c|c|c|c|}
\hline 
Dataset  & $n$  & $p$  & $n/p$  & $k$  & $hr$ \tabularnewline
\hline 
Iris  & 150  & 4  & 37.5  & 3  & 0.5\tabularnewline
Glass  & 214  & 9  & 23.7  & 6  & 0.25\tabularnewline
Wine  & 178  & 13  & 13.7  & 3  & 0.5\tabularnewline
Ionosphere  & 351  & 34  & 10.3  & 2  & 0.5\tabularnewline
Sonar  & 208  & 60  & 3.5 & 2  & 0.5 \tabularnewline
USPS 358  & 2248  & 256  & 8.8  & 3  & 0.5\tabularnewline
\hline 
\end{tabular}
\par\end{centering}

\caption{\label{tab:data}Data used in the experiments. $n$ is the number
of samples, $p$ is the number of features, $k$ is the number of
classes and $hr$ is the hold-out ratio used in the experiments.}
\end{table}

Each data set was randomly split into training and testing sets in
the hold-out ratio $hr$ given in Table~\ref{tab:data}. The data
were scaled between -1 and 1 on each variable. The search range for
the cross-validation was for the kernel hyperparameter $\sigma\in[-4,4]$,
for the common intrinsic dimension $d\in[1,20]$, for the scree test
threshold $\tau\in[10^{-7},1]$, for the regularization parameter
in KFD and KQDA $\lambda\in[10^{-13},10^{-6}]$ and for the penalty
parameter of the SVM $\gamma\in[2^{5},2^{9}]$. The global classification
accuracy was computed on the testing set and the reported results
have been averaged over 50 replications of the whole process. The
average classification accuracies and the standard deviations are
given in Table~\ref{tab:res_num}.

\begin{sidewaystable}
\begin{centering}
\begin{tabular}{|l||c|c|c|c|c|c||c|}
\hline 
Method  & Iris  & Glass  & Wine  & Ionosphere  & Sonar  & USPS 358  & Mean (rank)\tabularnewline
\hline 
pgpDA $\mathcal{M}_{0}$  & \textbf{95.9$\pm$ 2.1}  & 64.9 $\pm$ 6.3  & 96.8 $\pm$ 1.7  & 90.5 $\pm$ 2.3  & 77.9 $\pm$ .9  & 92.2 $\pm$ 1.0  & 86.4 (5) \tabularnewline
pgpDA $\mathcal{M}_{1}$  & 95.2$\pm$ 2.1  & 62.6 $\pm$ 12.5  & 96.7 $\pm$ 2.3  & \textbf{93.7 $\pm$ 1.6}  & \textbf{81.8 $\pm$ 4.9 }  & \cellcolor[gray]{0.9}\textbf{96.6 $\pm$ 0.4 }  & 87.8 (2)\tabularnewline
pgpDA $\mathcal{M}_{2}$  & 94.4$\pm$ 6.2  & 64.4 $\pm$ 6.7  & 96.8 $\pm$ 1.8  & 91.0 $\pm$ 2.8  & 71.6 $\pm$ 13.4  & 95.4 $\pm$ 0.8  & 85.6 (7)\tabularnewline
pgpDA $\mathcal{M}_{3}$  & 95.8$\pm$ 2.3  & 64.3 $\pm$ 6.8  & 96.9 $\pm$ 2.0  & 93.2 $\pm$ 2.1  & 79.3 $\pm$ 4.9  & 96.2 $\pm$ 0.5  & 87.6 (3)\tabularnewline
pgpDA $\mathcal{M}_{4}$  & 94.4$\pm$ 2.2  & \textbf{65.3 $\pm$ 6.4}  & \cellcolor[gray]{0.9}\textbf{97.2 $\pm$ 1.8}  & 93.4 $\pm$ 2.0  & 81.6 $\pm$ 4.5  & 96.3 $\pm$ 0.7  & \cellcolor[gray]{0.9}\textbf{88.0 (1)}\tabularnewline
pgpDA $\mathcal{M}_{5}$  & 94.2$\pm$ 7.1  & 59.8 $\pm$ 10.9  & 96.4 $\pm$ 2.0  & 92.0 $\pm$ 1.8  & 72.5 $\pm$ 12.6  & 96.0 $\pm$ 0.5 & 85.2 (8)\tabularnewline
pgpDA $\mathcal{M}_{6}$  & 94.8$\pm$ 2.1  & 65.2 $\pm$ 5.6  & \cellcolor[gray]{0.9}\textbf{97.2 $\pm$ 1.8}  & 92.5 $\pm$ 2.1  & 79.8 $\pm$ 4.9  & 96.1 $\pm$ 0.5 & 87.6 (3)\tabularnewline
pgpDA $\mathcal{M}_{7}$  & 41.3$\pm$ 16.5  & 40.0 $\pm$ 5.4  & 75.2 $\pm$ 8.3  & 64.6 $\pm$ 2.6  & 48.8 $\pm$ 5.7  & 63.5 $\pm$ 1.5 & 55.5 (11)\tabularnewline
pgpDA $\mathcal{M}_{8}$  & 29.2$\pm$ 17.4  & 35.4 $\pm$ 7.9  & 64.2 $\pm$ 26.8  & 64.3 $\pm$ 2.5  & 50.5 $\pm$ 5.5  & 36.8 $\pm$ 1.2 & 46.7 (12)\tabularnewline
\hline 
KFD  & 93.4$\pm$ 3.7  & 47.3 $\pm$ 10.1  & 95.9 $\pm$ 2.3  & \cellcolor[gray]{0.9}\textbf{94.1 $\pm$ 1.7}  & 82.9 $\pm$ 3.1  & \textbf{93.6 $\pm$ 0.5 } & 84.5 (9)\tabularnewline
KQDA  & \cellcolor[gray]{0.9}\textbf{96.6$\pm$ 2.3}  & 64.5 $\pm$ 6.3  & 96.6 $\pm$ 1.7  & 88.1 $\pm$ 2.3  & 68.9$\pm$18.1  & 64.7 $\pm$ 37.5 & 79.9 (10)\tabularnewline
SVM  & 95.7$\pm$ 2.0  & \cellcolor[gray]{0.9}\textbf{69.1 $\pm$ 5.5 }  & \textbf{96.8 $\pm$ 1.4}  & 92.8 $\pm$ 1.8  & \cellcolor[gray]{0.9}\textbf{84.8 $\pm$ 4.0 }  & 77.6 $\pm$ 5.4 & \textbf{86.1 (6)}\tabularnewline
\hline 
\end{tabular}{\footnotesize{} \caption{\label{tab:res_num}Classification results on real-world datasets:
reported values are average correct classification rates and standard
deviation computed on validation sets.}
}
\par\end{centering}{\footnotesize \par}

\end{sidewaystable}

Regarding the competitive methods, KFD and SVM provide often better
results than KQDA. The model used in KQDA only fits ``ionosphere'',
{}''iris'' and ``wine'' data, for which classification accuracies
are similar to or better than those obtain with KFD and SVM. For the
parsimonious pgpDA models, except for $\mathcal{M}_{7}$ and $\mathcal{M}_{8}$,
the classification accuracies are globally good. Models $\mathcal{M}_{1}$
and $\mathcal{M}_{4}$ provide the best results in terms of average
correct classification rates. In particular, for the ``USPS 358''
and ``wine'' data sets, they provide the best overall accuracies.
Let us remark that pgpDA performs significantly better than SVM (for
the Gaussian kernel) on high-dimensional data (USPS~358).

In conclusion of these experiments, by relying on parsimonious models
rather than regularization, pgpDA provides good classification accuracies
and it is robust to the situation where few samples are available
in regards to the number of variables in the original space. In practice,
models $\mathcal{M}_{1}$ and $\mathcal{M}_{4}$ should be recommended:
intrinsic dimension is common between the classes and the variance
inside the class intrinsic subspace is either free or common. Conversely,
models $\mathcal{M}_{7}$ and $\mathcal{M}_{8}$ must be avoided since
they appeared to be too constrained to handle real classification
situations.

\subsection{Classification of functional data: the Canadian temperatures}

\begin{figure}
\begin{centering}
\begin{tabular}{cc}
\includegraphics[bb=25bp 40bp 475bp 455bp,clip,width=0.48\columnwidth]{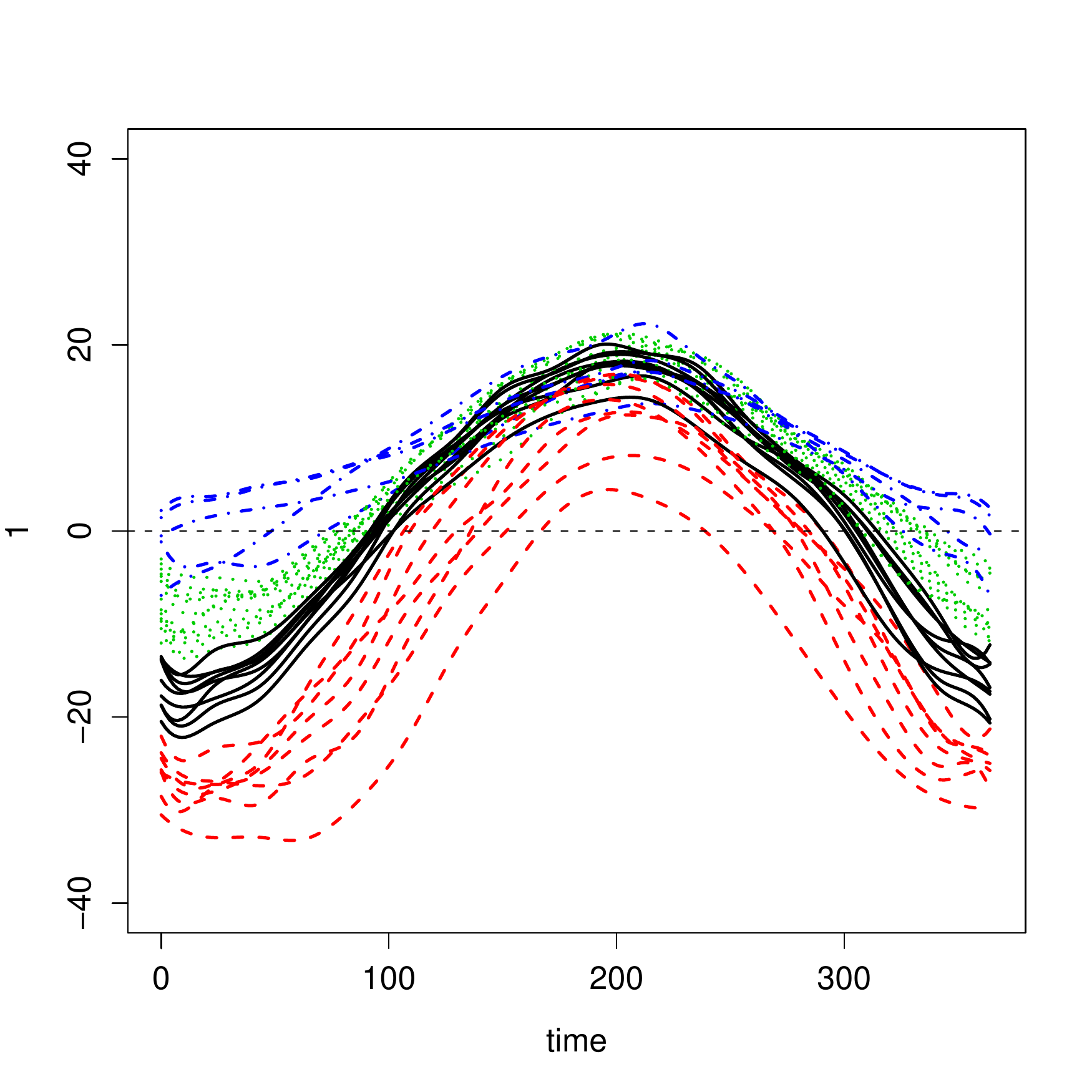}  & \includegraphics[bb=55bp 55bp 450bp 450bp,clip,width=0.48\columnwidth,height=0.28\textheight]{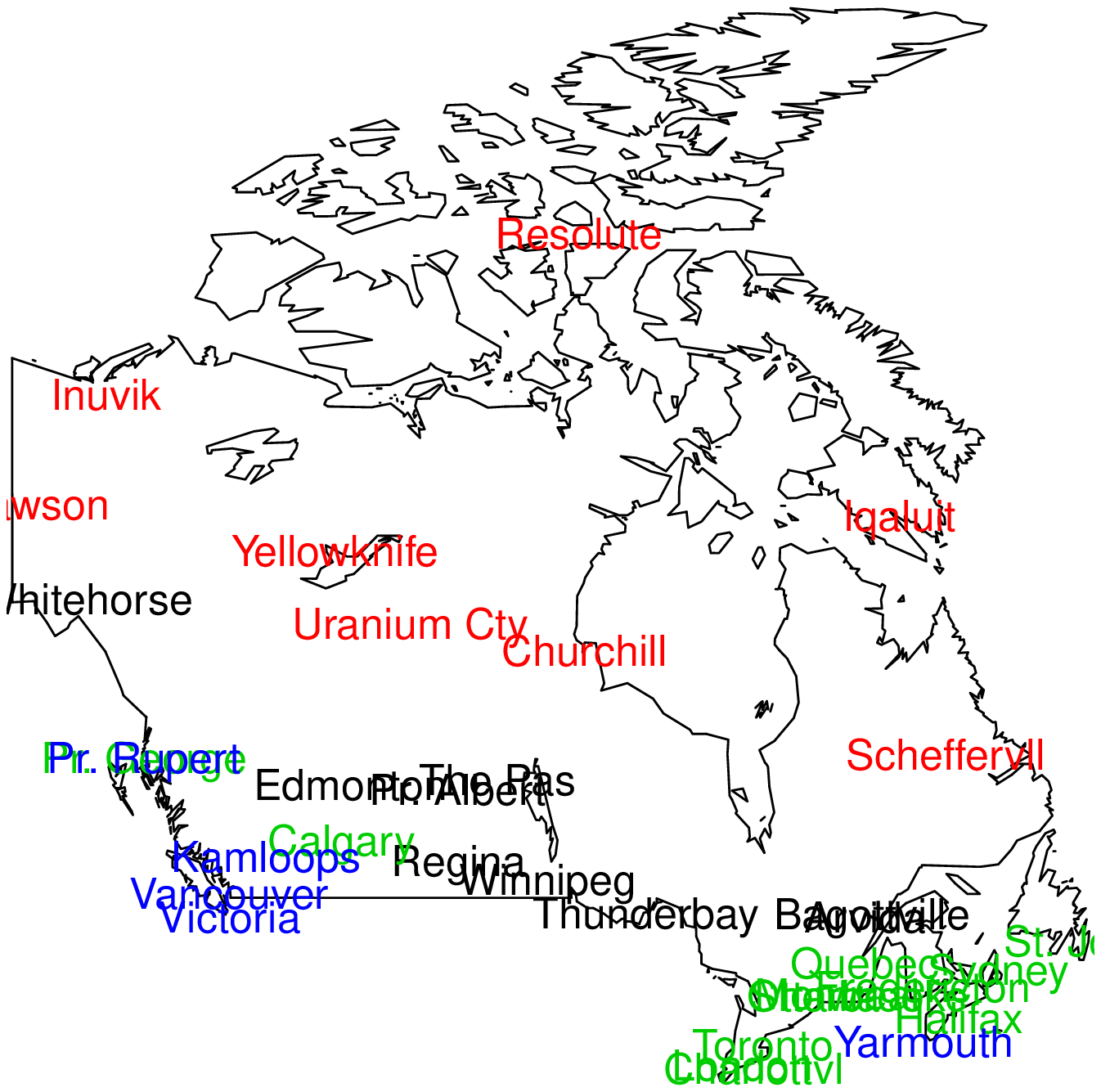} \tabularnewline
\end{tabular}
\par\end{centering}

\caption{\label{Canada-1}Clustering of the 35 times series of the Canadian
temperature data set into 4 groups with pgpEM (left) and geographical
positions of the weather stations according to their group belonging
(right). The colors indicate the group memberships: group 1 in black,
\textcolor{red}{group 2 in red}, \textcolor{green}{group 3 in green}
and \textcolor{blue}{group 4 in blue}.}
\end{figure}

We now focus on illustrating the possible range of application of
the proposed methodologies to different types of data. We consider
here the clustering of functional data with pgpEM for which the mapping
function $\varphi$ is explicit (see Section~\ref{sub:Case-of-functional}).
The Canadian temperature data used in this study, presented in details
in~\cite{Ram2005}, consist in the daily measured temperatures at
35 Canadian weather stations across the country. The pgpEM algorithm
was applied here with the model $\mathcal{M}_{0}$, which is the most
general parsimonious Gaussian process model proposed in this work,
with a fixed number $k$ of groups set to $4$. The mapping function
$\varphi$ consists in the projection of the observed curves on a
basis of 20 natural cubic splines. Once the pgpEM algorithm has converged,
various informations are available and some of them are of particular
interest. Group means, intrinsic dimensions of the group-specific
subspaces and functional principal components of each group could
in particular help the practitioner in understanding the clustering
of the dataset at hand. The left panel of Figure~\ref{Canada-1}
presents the clustering of the temperature data set into 4 groups
with pgpEM.

It is first interesting to have a look at the name of the weather
stations gathered in the different groups formed by pgpEM. It appears
that group~1 (black solid curves) is mostly made of continental stations,
group~2 (red dashed curves) mostly gathers the stations of the North
of Canada, group~3 (green dotted curves) mostly contains the stations
of the Atlantic coast whereas the Pacific stations are mostly gathered
in group~4 (blue dot-dashed curves). For instance, group~3 contains
stations such as Halifax (Nova Scotia) and St Johns (Newfoundland)
whereas group~4 has stations such as Vancouver and Victoria (both
in British Columbia). The right panel of Figure~\ref{Canada-1} provides
a map of the weather stations where the colors indicate their group
membership. This figure shows that the obtained clustering with pgpEM
is very satisfying and rather coherent with the actual geographical
positions of the stations (the clustering accuracy is 71\% here compared
with the geographical classification provided by \cite{Ram2005}).
We recall that the geographical positions of the stations have not
been used by pgpEM to provide the partition into 4 groups.

\begin{figure}[p]

\begin{centering}
\begin{tabular}{c}
\includegraphics[width=0.95\columnwidth]{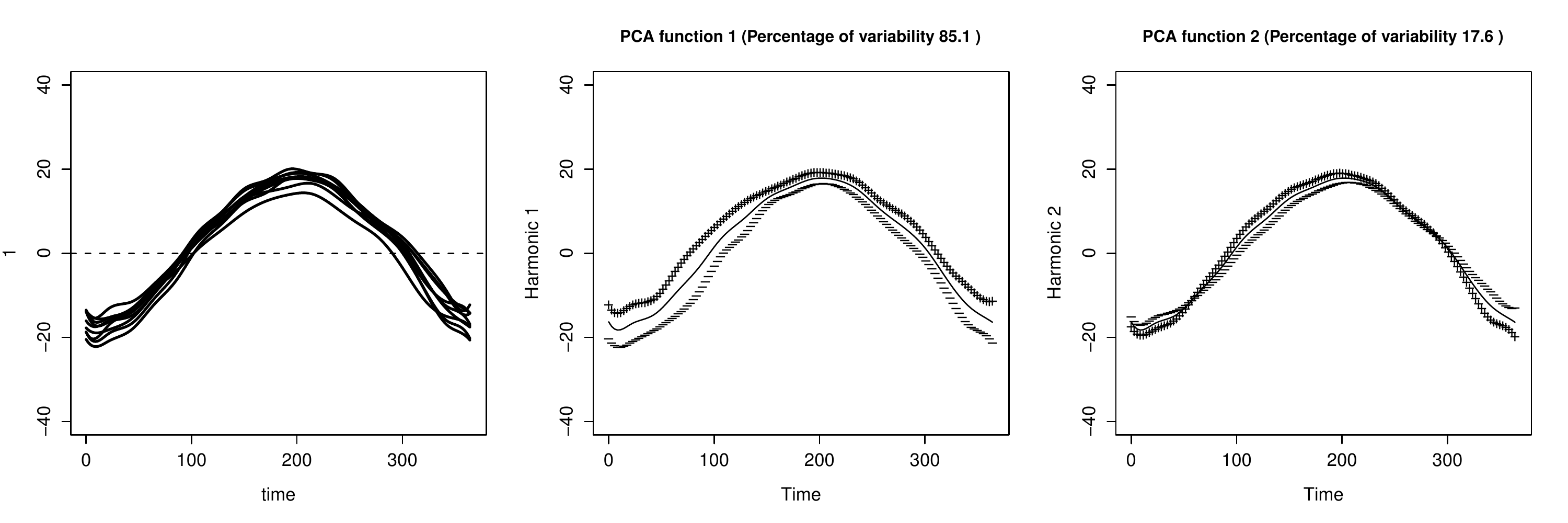}\tabularnewline
{\small (a) Group~1 (mostly continental stations)}\tabularnewline
\includegraphics[width=0.95\columnwidth]{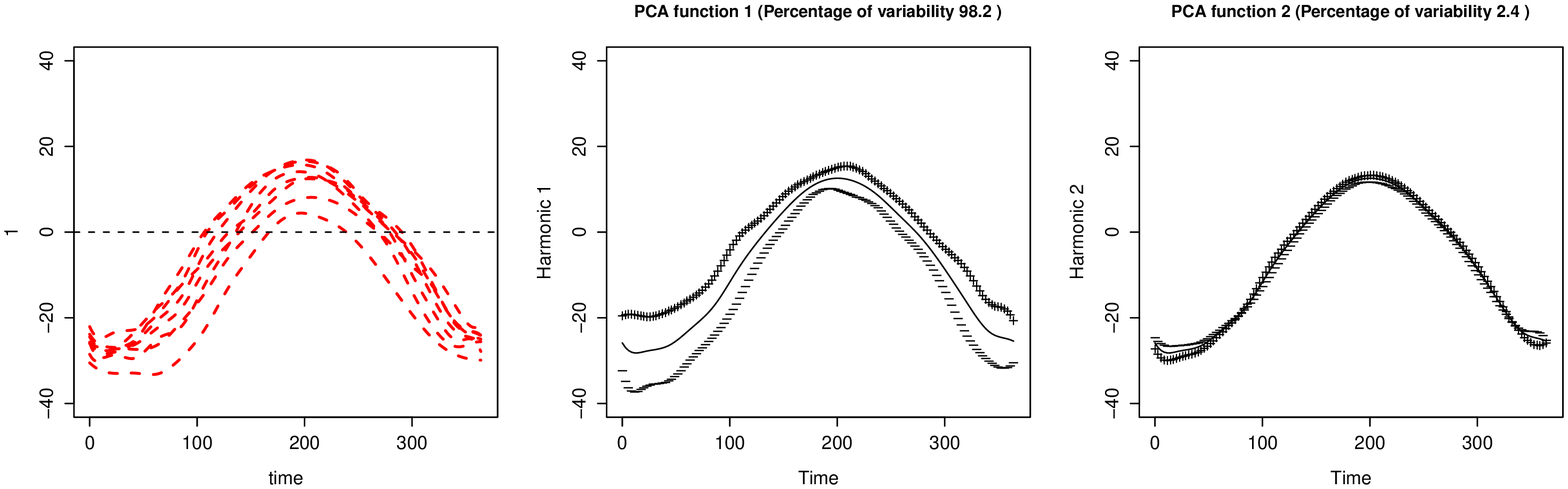}\tabularnewline
{\small (b) Group~2 (mostly Arctic stations)}\tabularnewline
\includegraphics[width=0.95\columnwidth]{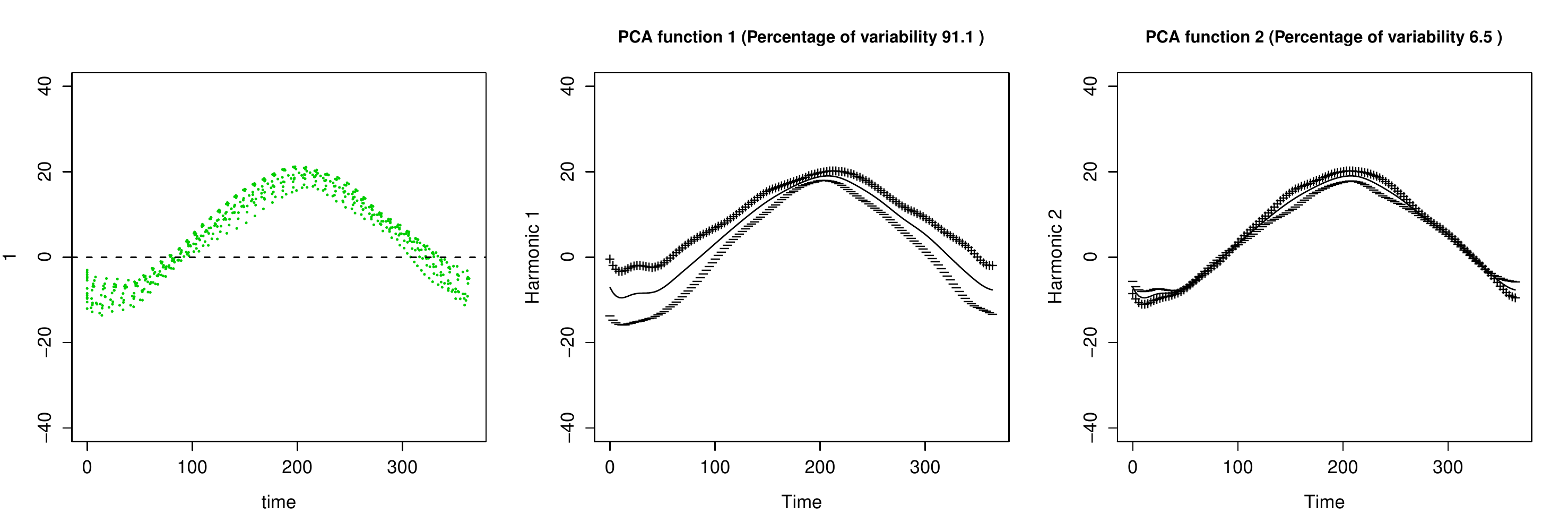}\tabularnewline
{\small (c) Group~3 (mostly Atlantic stations)}\tabularnewline
\includegraphics[width=0.95\columnwidth]{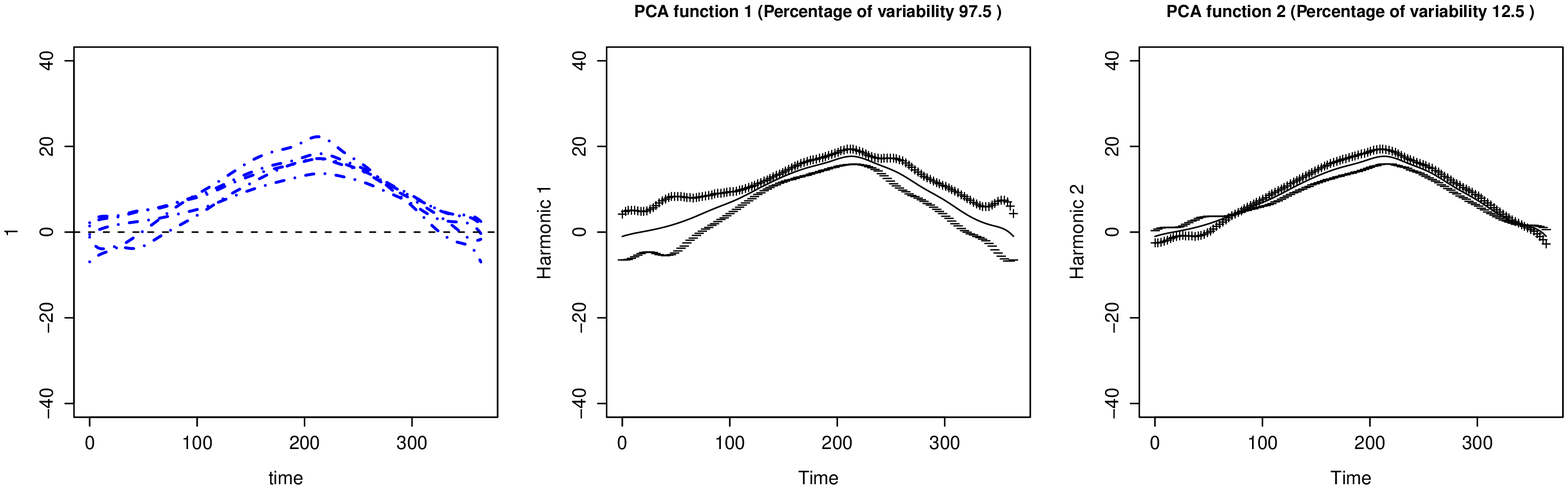}\tabularnewline
{\small (d) Group~4 (mostly Pacific stations)}\tabularnewline
\end{tabular}
\par\end{centering}

\caption{\label{Canada-3}The group means of the Canadian temperature data
obtained with pgpEM and the effects of adding (+) and subtracting
(−) twice the square root of the feature subspace variance (see text
for details).}
\end{figure}

An important characteristic of the groups, but not necessarily easy
to visualize, is the specific functional subspace of each group. A
classical way to observe principal component functions is to plot
the group mean function $\hat{\mu}_{i}(t)$ as well as the functions
$\hat{\mu}_{i}(t)\pm2\sqrt{\hat{\lambda}_{ij}}\hat{q}_{ij}(t)$ (see~\cite{Ram2005}
for more details). Figure~\ref{Canada-3} shows such a plot for the
4 groups of weather stations formed by pgpEM. It first appears on
the first functional principal component of each group that there
is more variance between the weather stations in winter than in summer.
In particular, the first principal component of group~4 (blue curves,
mostly Pacific stations) reveals a specific phenomenon which occurs
at the beginning and the end of the winter. Indeed, we can observe
a high variance in the temperatures of the Pacific coast stations
at these periods of time which can be explained by the presence of
mountain stations in this group. The analysis of the second principal
components reveals finer phenomena. For instance, the second principal
component of group~1 (black curves, mostly continental stations)
shows a slight shift between the + and − along the year which indicates
a time-shift effect. This may mean that some cities of this group
have their seasons shifted, \textit{e.g.} late entry and exit in the
winter. Similarly, the inversion of the + and − on the second principal
component of the Pacific and Atlantic groups (blue and green curves)
suggests that, for these groups, the coldest cities in winter are
also the warmest cities in summer. On the second principal component
of group~2 (red curves, mostly Arctic stations), the fact that the
+ and − curves are almost superimposed shows that the North stations
have very similar temperature variations (different temperature means
but same amplitude) along the year.

\subsection{Classification of networks: the Add Health dataset}

\begin{figure}
\begin{centering}
\includegraphics[width=0.6\columnwidth]{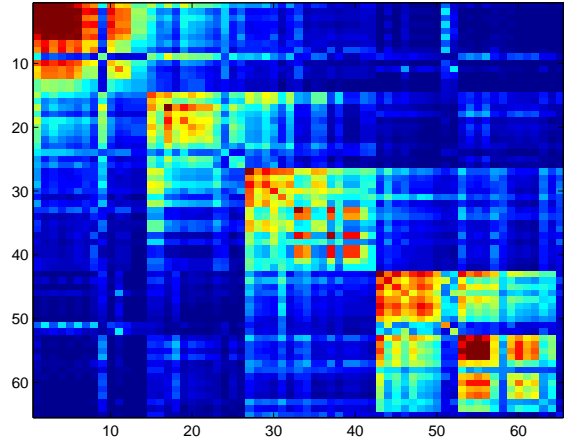} 
\par\end{centering}

\caption{\label{fig:Network-0}Regularized Laplacian kernel associated to the
Add Health network for $\nu=4$: blue pixels correspond to low values
(low similarity between nodes) and red pixels correspond to high values
(high similarity between nodes).}
\end{figure}

We now consider network data which are nowadays widely used to represent
relationships between persons in organizations or communities. Recently,
the need of classifying and visualizing such data has suddenly grown
due to the emergence of Internet and of a large number of social network
websites. Indeed, increasingly, it is becoming possible to observe
“network informations” in a variety of contexts, such as email transactions,
connectivity of web pages, protein-protein interactions and social
networking. A number of scientific goals can apply to such networks,
ranging from unsupervised problems such as describing network structure,
to supervised problems such as predicting node labels with information
on their relationships.

We investigate here the use of pgpDA to classify the nodes of a network.
To our knowledge, only a few kernels~(see~\cite{Smola09} for more
details) have been proposed for network data and the regularized Laplacian
kernel is probably the most used. This kernel is defined as follows:
let $X$ be a symmetric $n\times n$ socio-matrix where $X_{ij}=1$
if a relationship is observed between the nodes $i$ and $j$ and
$X_{ij}=0$ in the opposite case. Let $D$ be the diagonal matrix
where $D_{ii}$ indicates the number of relationships for the node
$i$, \emph{i.e., $D_{ii}=\sum_{j=1}^{n}X_{ij}$}. The regularized
Laplacian kernel $K$ is then defined by: 
\[
K=\left[\tilde{L}+\nu I_{n}\right]^{-1},
\]
 where $\tilde{L}=I_{n}-D^{-\frac{1}{2}}XD^{-\frac{1}{2}}$ is the
normalized Laplacian of the network, $\nu$ is a positive value and
$I_{n}$ is the identity matrix of size $n$.

The social network studied here is from the National Longitudinal
Study of Adolescent Health and it is a part of a big dataset, usually
called the “Add Health” dataset. The data were collected in 1994-95
within 80 high-schools and 52 middle schools in the USA. The whole
study is detailed in~\cite{Harris03}. In addition to personal and
social information, each student was asked to nominate his best friends.
We consider here the social network based on the answers of 67 students
from a single school, treating the grade of each student as the class
variable. Two adolescents who nominated nobody were removed from the
network. We therefore consider a whole dataset made of 65 students
distributed into 5 classes: grade~7 to grade~11.

\begin{figure}
\begin{centering}
\begin{tabular}{cc}
\includegraphics[bb=90bp 265bp 495bp 575bp,clip,width=0.48\columnwidth]{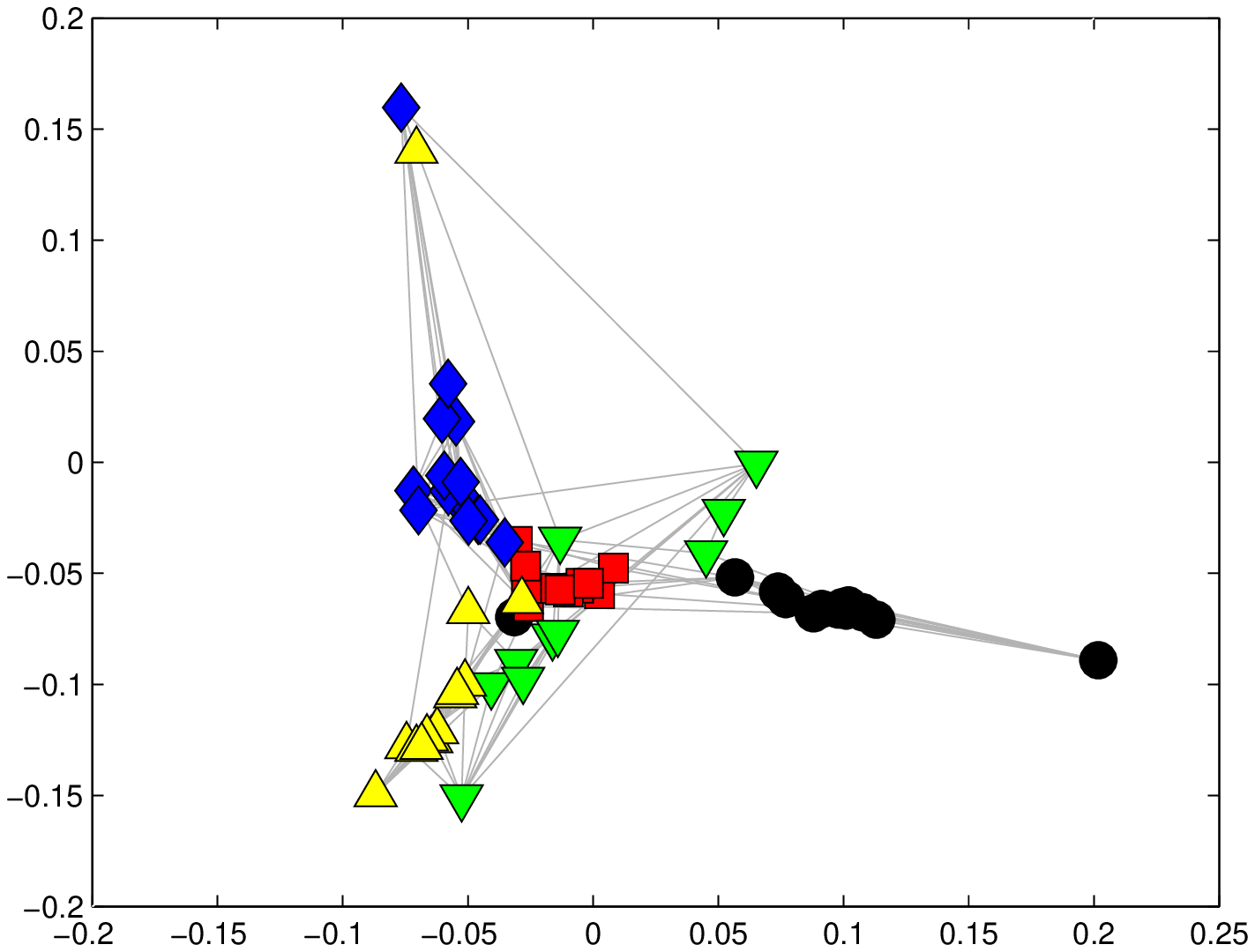}  & \includegraphics[bb=90bp 265bp 495bp 575bp,clip,width=0.48\columnwidth]{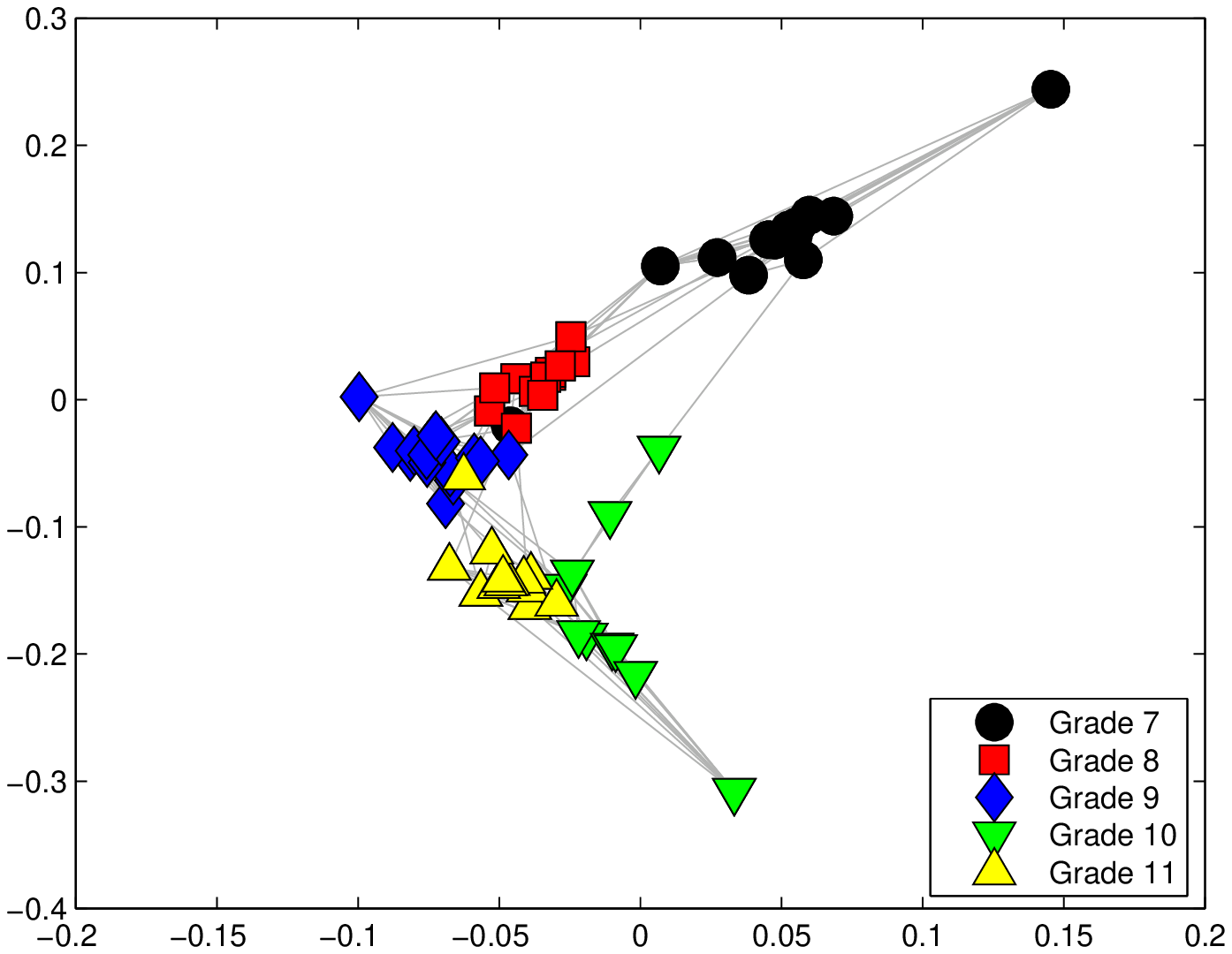}\tabularnewline
(a) Subspace of class 2  & (b) Subspace of class 4\tabularnewline
\end{tabular}
\par\end{centering}

\caption{\label{fig:Network-2}Visualisation of the Add Health network with
pgpDA in the feature subspace of the 2nd and the $4$th class (grade
8 and 10 respectively).}
\end{figure}

We first selected by cross-validation the kernel parameter on a learning
sample and the threshold parameter for the intrinsic dimensions was
set to $0.2$. The most adapted value for $\nu$ was 4 and this gives
on average 96.92\% of correct classification for the test nodes. Remark
that $\nu$ turned out not to be a sensitive parameter and we obtain
satisfying results for a large range of values of $\nu$. Figure~\ref{fig:Network-0}
presents the kernel associated with the selected value of $\nu$.
Since network visualization is an important issue in network analysis,
we then kept these parameters to visualize the whole network in the
feature subspace of each class. Figure~\ref{fig:Network-2} presents
the visualization of the network into the feature subspace of the
classes 2 and 4. Both visualizations turn out to be very informative
and, in particular, the visualization on the feature subspace of the
4th class (grade 10) is particularly useful to understand the network.
It is interesting to notice that the network is almost organized along
a 1-dimensional manifold (an half-circle here) which is consistent
with the nature of the network: students of different classes. The
specific form of the representation is due here to some relations
between students of grade 7 and 10 (students of the same family perhaps).
We also remark that the classes are quite well separated and most
of the relationships between students of different classes are between
consecutive grades. This suggests that relationships between classes
are due to students who failed to move to the upper grade and who
may keep contact with old friends. It is in addition interesting to
notice that this visualization is very close to the one obtained on
the same network by Hoff, Handcock and Raftery in~\cite{Handcock07}
using the so-called ``latent space model''.

\subsection{Classification of categoretical data: the house-vote dataset}

\begin{figure}
\begin{centering}
\includegraphics[width=1\columnwidth]{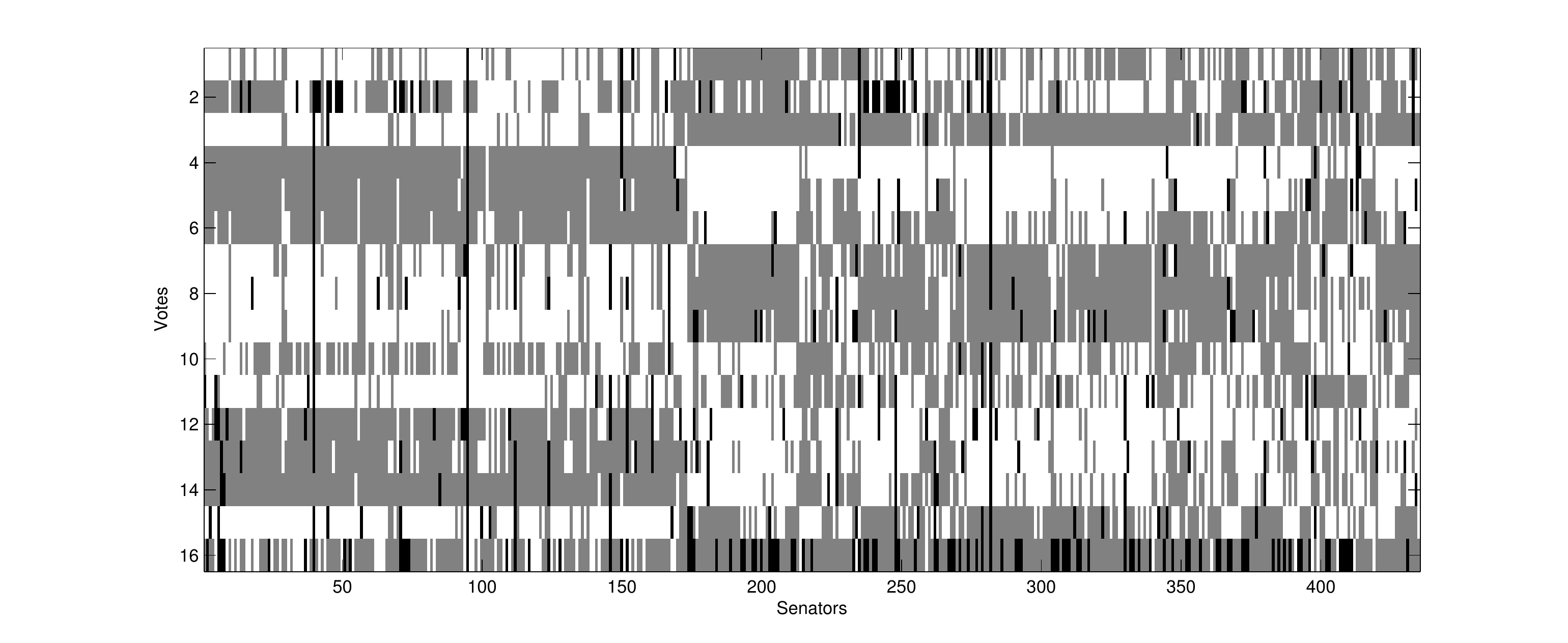} 
\par\end{centering}

\caption{\label{fig:Votes-1}Votes (yea, nay or unknown) for each of the U.S.
House of Representatives congressmen on 16 key votes in 1984. Yeas
are in indicated in white, nays in gray and missing values in black.
The first 168 congressmen are republicans whereas the 267 last ones
are democrats.}
\end{figure}

We focus now on categorical data which are also very frequent in scientific
fields. We consider here the task of clustering (unsupervised classification)
and therefore the pgpEM algorithm. To evaluate the ability of pgpEM
to classify categorical data, we used the U.S. House Votes data set
from the UCI repository. This data set is a record of the votes (yea,
nay or unknown) for each of the U.S. House of Representatives congressmen
on 16 key votes in 1984. These data were recorded during the during
the third and fourth years of Ronald Reagan's Presidency. At this
time, the republicans controlled the Senate, while the democrats controlled
the House of Representatives. Figure~\ref{fig:Votes-1} shows the
database where yeas are in indicated in white, nays in gray and missing
values in black. The first 168 congressmen are republicans whereas
the 267 last ones are democrats. As we can see, the considered votes
are very discriminative since republicans and democrats vote differently
in almost all cases while most of the congressmen follow the majority
vote in their group. We can however notice that a significant part
(around 50 congressmen) of the democrats tend to vote differently
from the other democrats.

\begin{figure}
\begin{centering}
\begin{tabular}{cc}
\includegraphics[bb=110bp 275bp 480bp 570bp,clip,width=0.47\columnwidth]{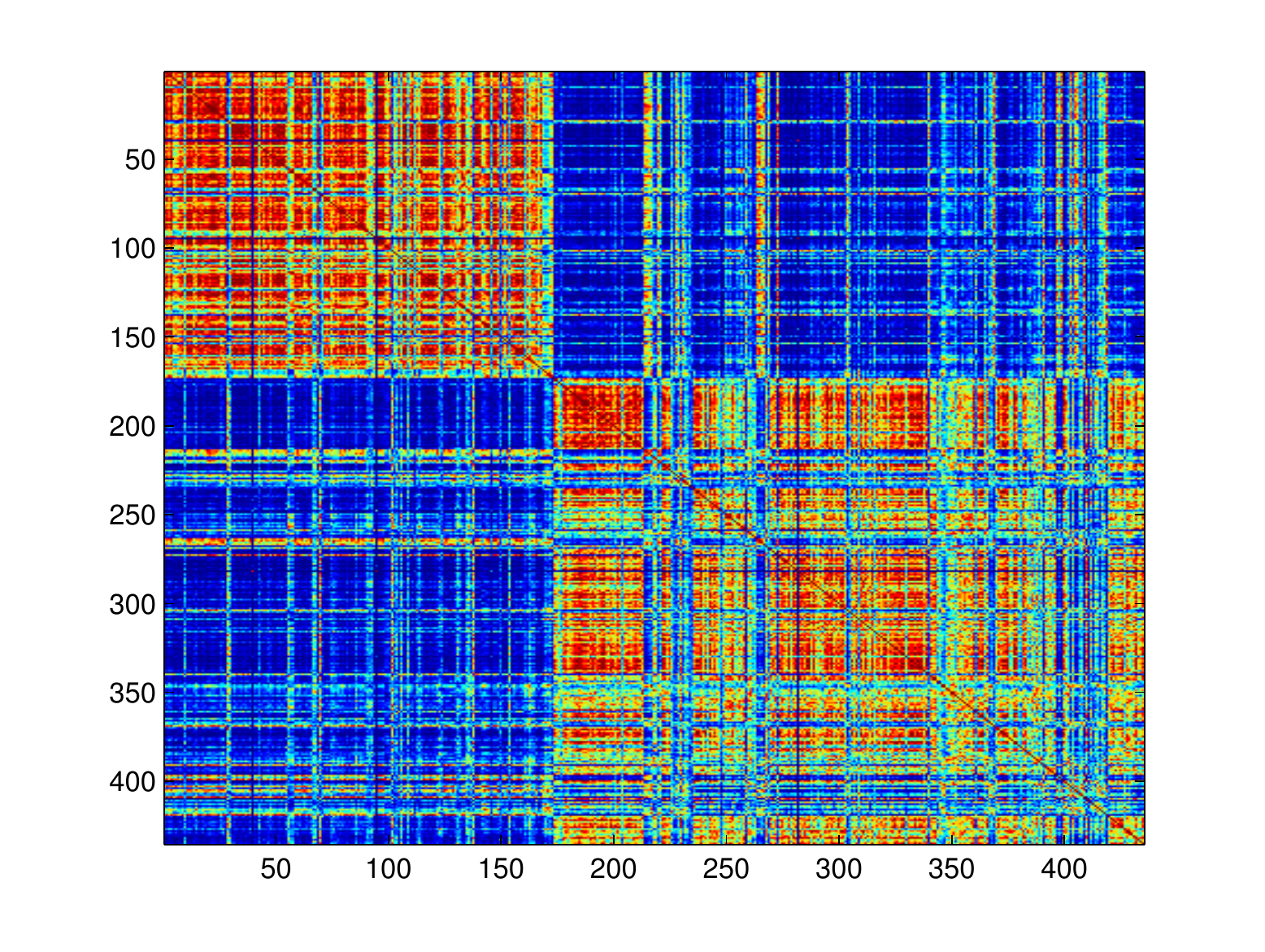}  & \includegraphics[bb=110bp 275bp 480bp 570bp,clip,width=0.47\columnwidth]{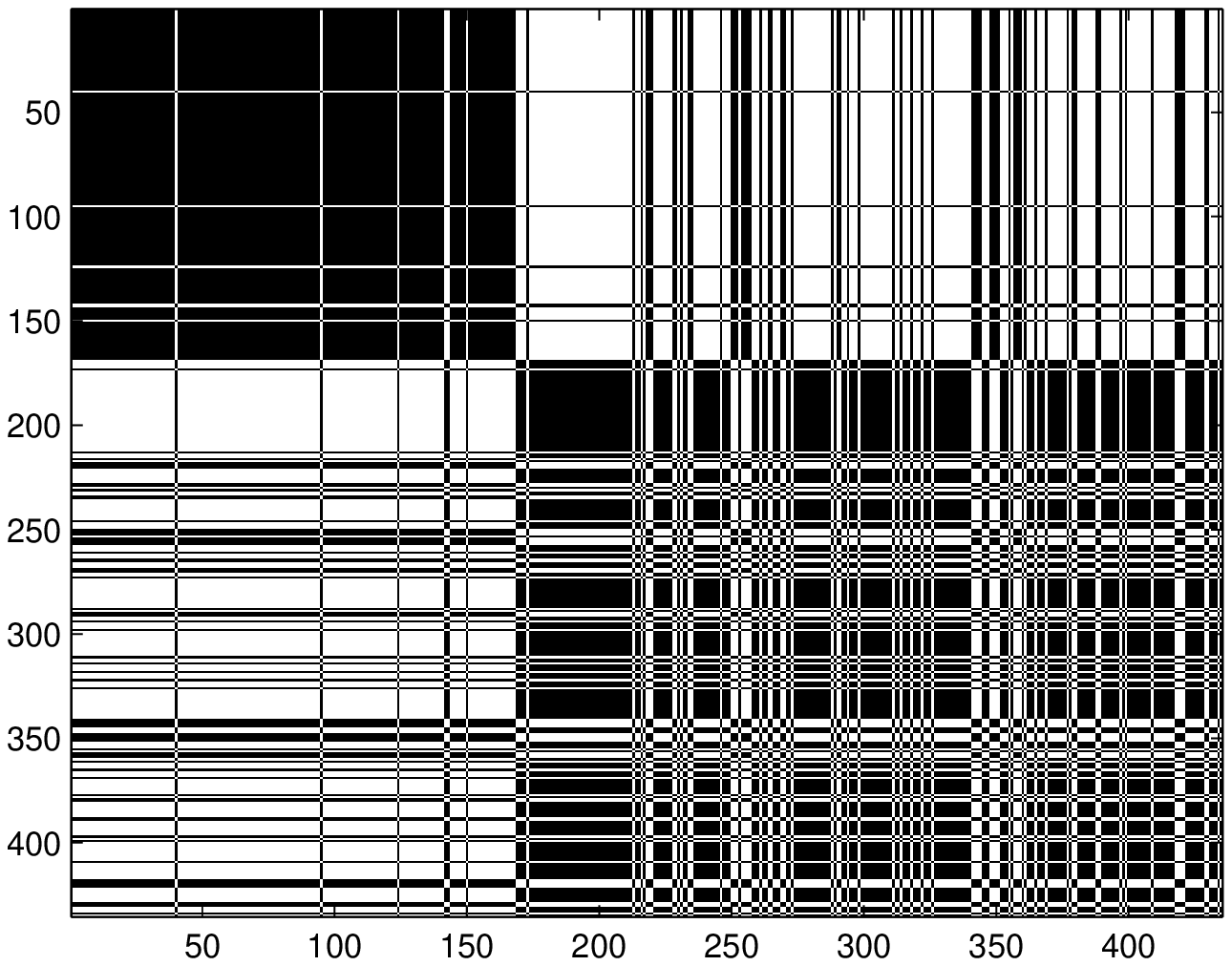}\tabularnewline
\end{tabular}
\par\end{centering}

\caption{\label{fig:Votes-2}Kernel based on the Hamming distance (left) computed
on the house-vote dataset and clustering results (right) obtained
with pgpEM. For the kernel, blue and red pixels correspond respectively
to low and high values. The clustering results are presented through
a binary matrix where a black pixel indicates a common membership
between two senators.}
\end{figure}

\begin{figure}
\begin{centering}
\begin{tabular}{cc}
\includegraphics[bb=100bp 270bp 485bp 570bp,clip,width=0.47\columnwidth]{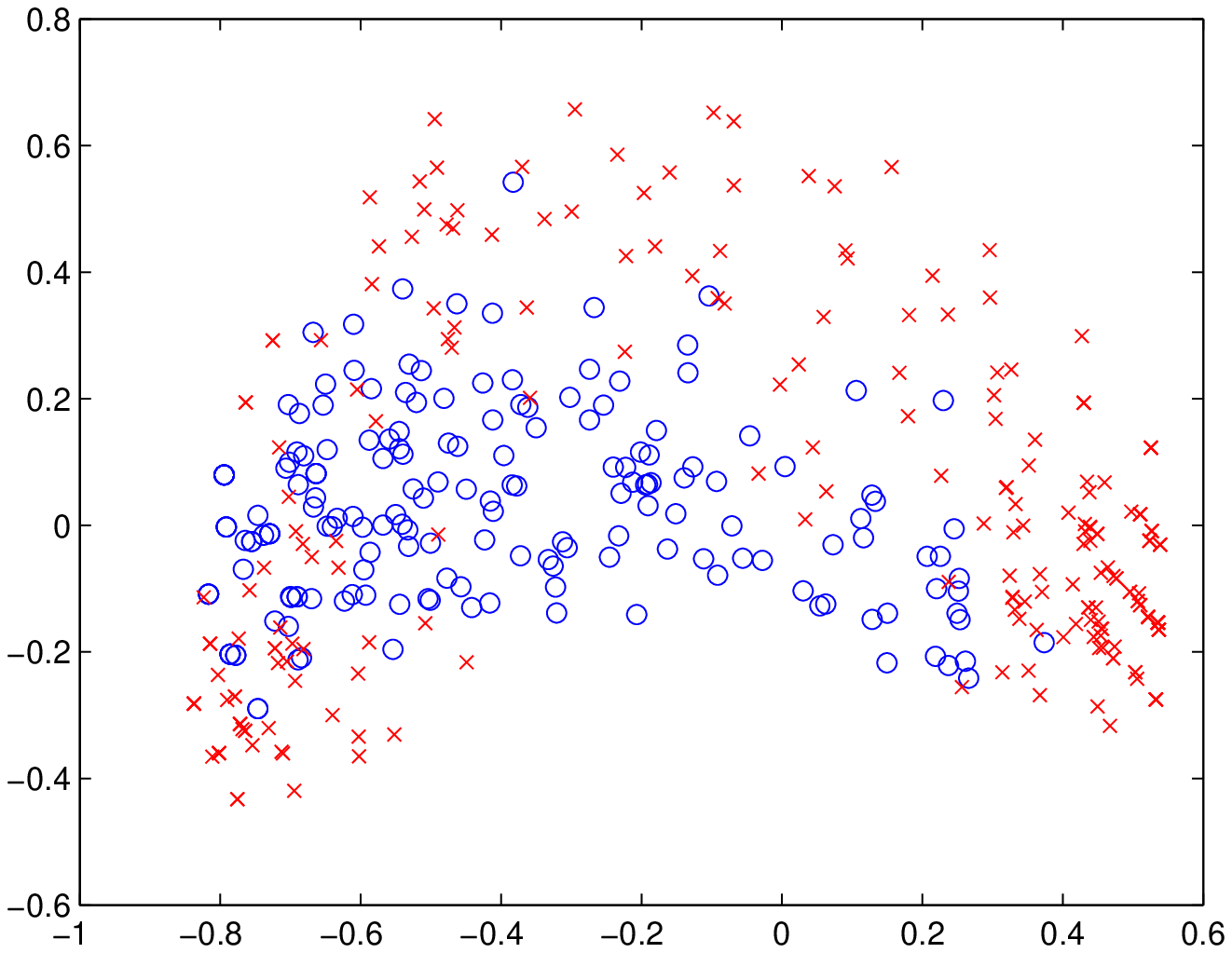}  & \includegraphics[bb=100bp 270bp 485bp 570bp,clip,width=0.47\columnwidth]{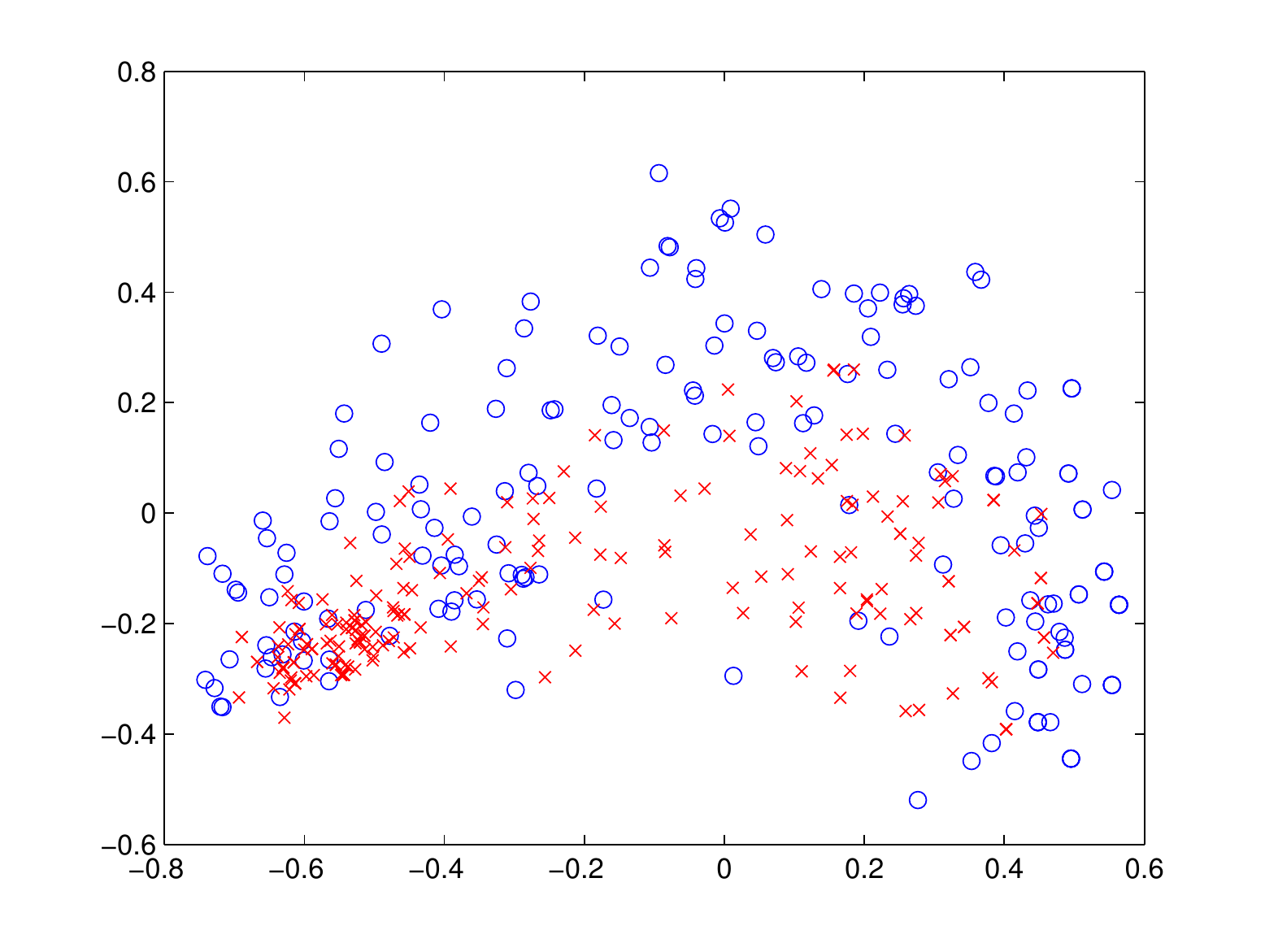}\tabularnewline
\end{tabular}
\par\end{centering}

\caption{\label{fig:Votes-3}Visualization of the house-vote data in the feature
subspace of the republican (left) and the democrat (right) groups
(red crosses denote republicans and blue circles denote democrats).}
\end{figure}

To cluster this dataset, we first build a kernel from the categorical
observations (16 qualitative variables with 3 possible values: yea,
nay or ?). We chose a kernel, proposed in~\cite{Couto05}, based
on the Hamming distance which measures the minimum number of substitutions
required to change one observation into another one. Figure~\ref{fig:Votes-2}
presents the resulting kernel (left panel) and the clustering result
obtained with the pgpEM algorithm. The clustering results are presented
through a binary matrix where a black pixel indicates a common membership
between two senators and a white pixel means different memberships
for the two senators. The pgpEM algorithm was used with the model
$\mathcal{M}_{0}$, with a number of group equals to 2 and the Cattell's
threshold was set to 0.2. The clustering accuracy between the obtained
partition of the data and the democrat/republican partition was 84.37\%
on this example. As one can observe, the pgpEM algorithm globally
succeeds in recovering the partition of the House of Representatives.
It is also interesting to notice that most of the congressmen which
are not correctly classified are those who tend to vote differently
from the majority vote in their group. Finally, the pgpEM algorithm
allows to visualize the observed categorical data into the (quantitative)
feature subspace of the two groups. Figure~\ref{fig:Votes-3} presents
these visualizations. The observation of these two plots confirms
the fact that republicans voted more homogeneously than democrats
in 1984 since there is no clear concentration of points on both plots
for the democrats.

\subsection{Classification of mixed data: the Thyroid dataset}

In this final experiment, we consider the supervised classification
of mixed data which is more and more a frequent case. Indeed, it is
usual to collect for the same individuals both quantitative and categorical
data. For instance, in Medicine, several quantitative features can
be measured for a patient (blood test results, blood pressure, morphological
characteristics, ...) and these data can be completed by answers of
the patient on its general health conditions (pregnancy, surgery,
tabacco, ...). The Thyroid dataset considered here is from the UCI
repository and contains thyroid disease records supplied by the Garavan
Institute, Sydney, Australia. The dataset contains 665 records on
male patients for which the answers (true of false) on 14 questions
have been collected as well as 6 blood test results (quantitative
measures). Among the 665 patients of the study, 61 suffer from a thyroid
disease.

To make pgpDA able to deal with such data, we built a combined kernel
by mixing a kernel based on the Hamming distance~\cite{Couto05}
(same kernel as in the previous section) for the categorical features
and a Gaussian kernel for the quantitative data. We chose to combine
both kernels simply as follows: 
\[
K(x_{j},x_{\ell})=\alpha K_{1}(x_{j},x_{\ell})+(1-\alpha)K_{2}(x_{j},x_{\ell}),
\]
where $K_{1}$ and $K_{2}$ are the kernels computed respectively
on the categorical and quantitative features. Another solution would
be to multiply both kernels. We refer to~\cite{Mehmet_2011} for
further details on multiple kernel learning.

\begin{figure}
\begin{centering}
\begin{tabular}{ccc}
\includegraphics[bb=100bp 265bp 495bp 580bp,clip,width=0.31\columnwidth]{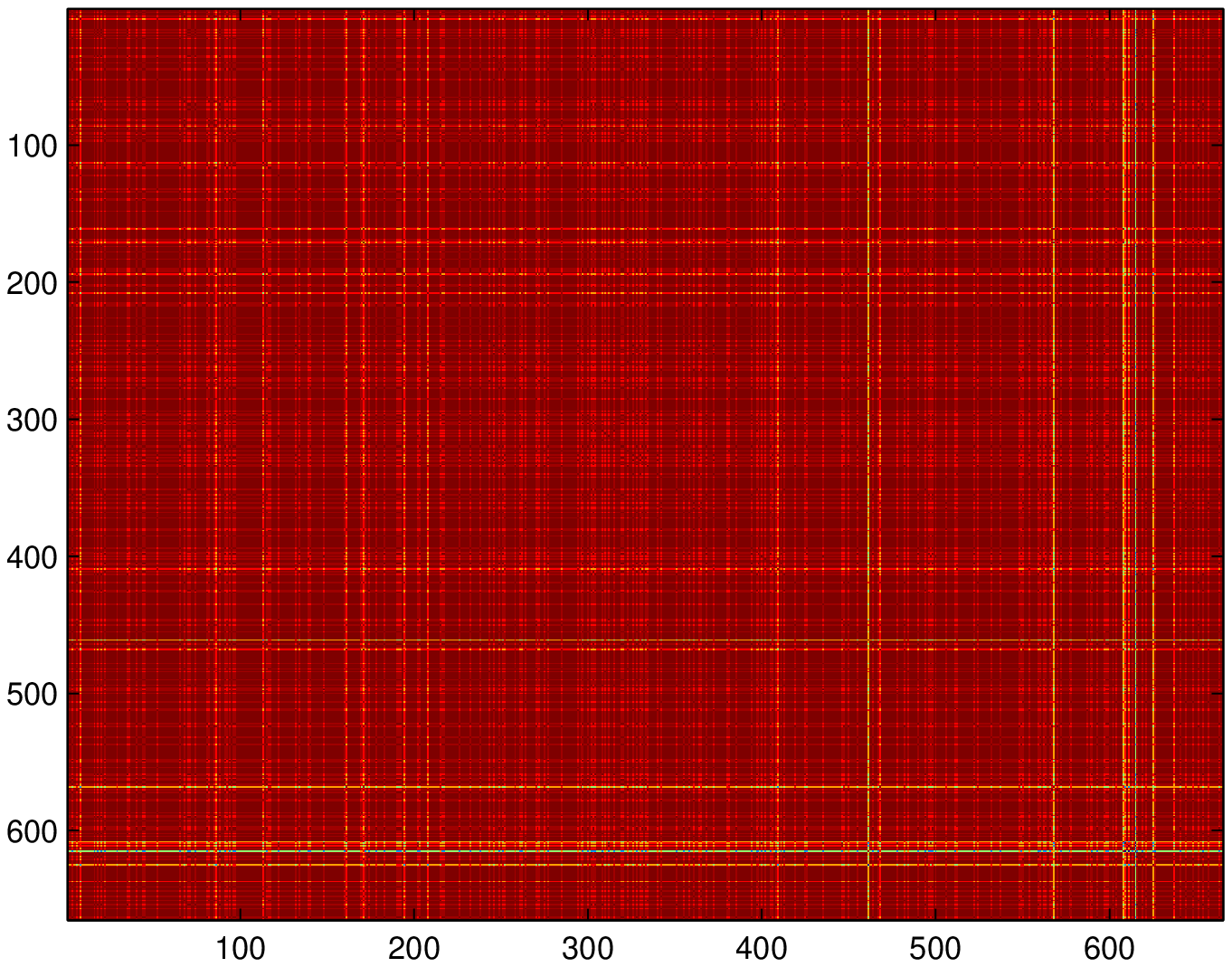}  & \includegraphics[bb=100bp 265bp 495bp 580bp,clip,width=0.31\columnwidth]{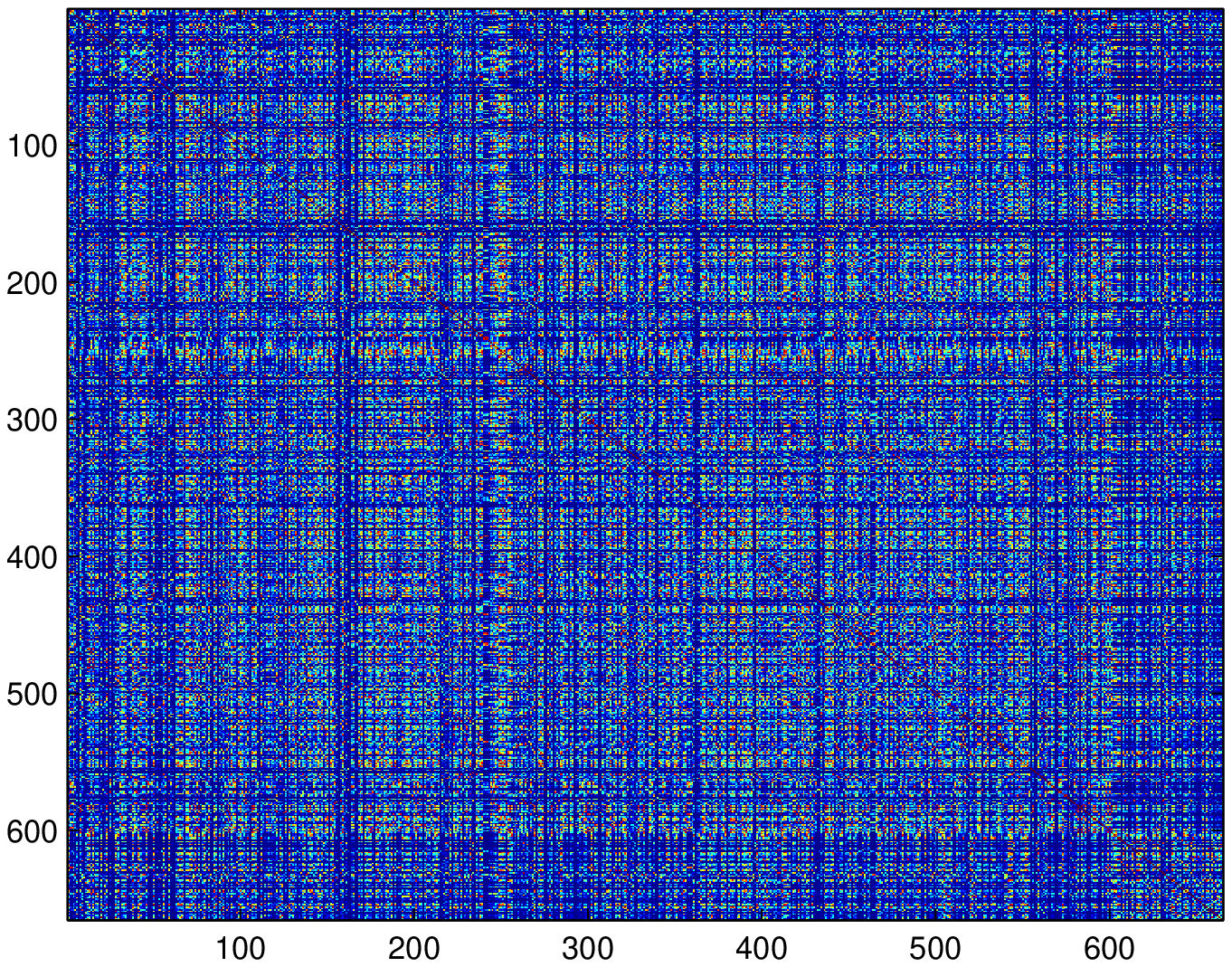}  & \includegraphics[bb=100bp 265bp 495bp 580bp,clip,width=0.31\columnwidth]{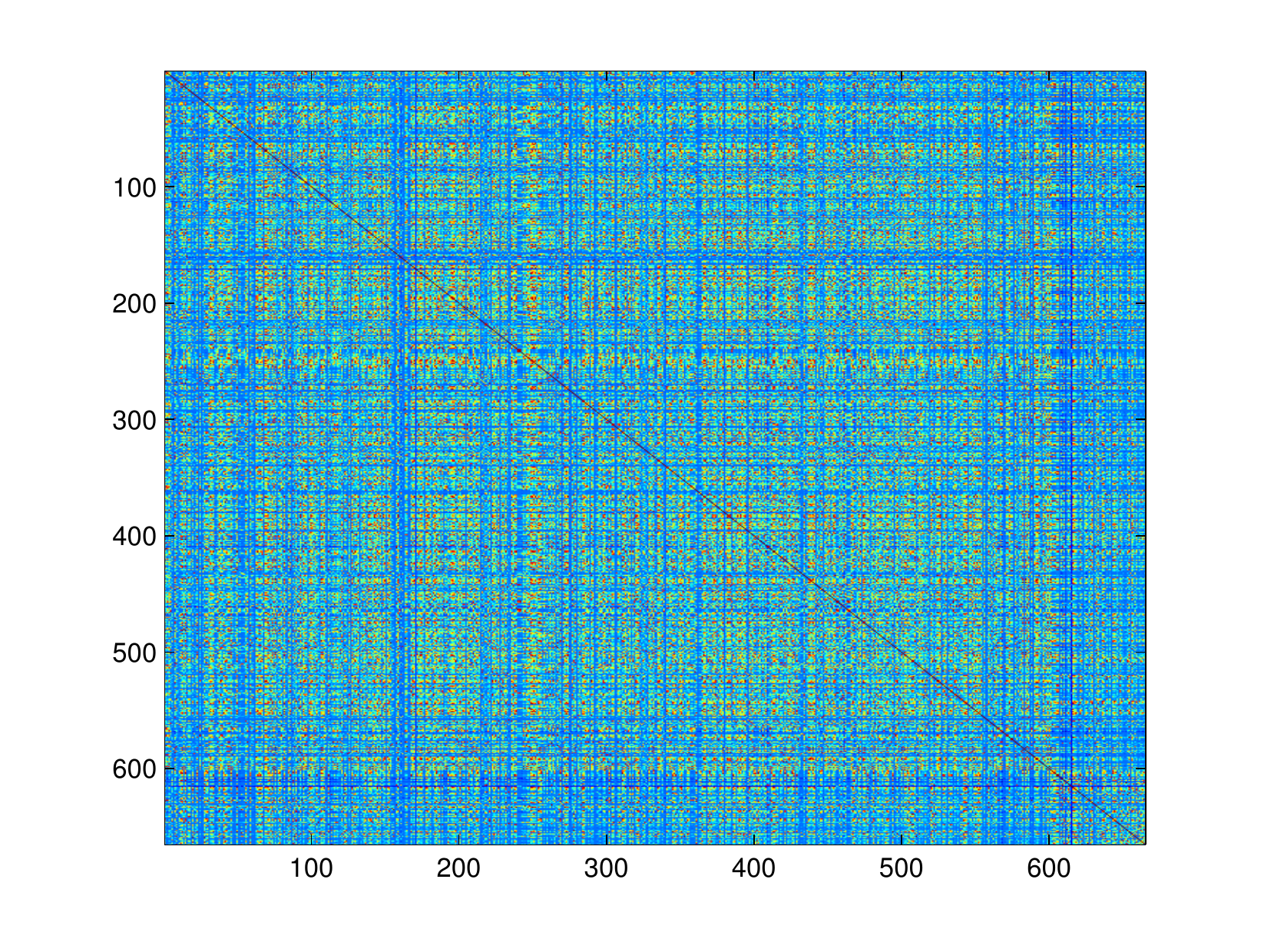}\tabularnewline
Quantitative data kernel  & Categorical data kernel  & Combined kernel\tabularnewline
\end{tabular}
\par\end{centering}

\caption{\label{fig:Mixt-2}Quantitative (left) and categorical (center) data
kernels and the combined kernel (right) for the Thyroid dataset (mixed
data). }
\end{figure}

We selected the optimal set of kernel parameters by cross-validation
on a learning part of the data. The model for pgpDA was the model
$\mathcal{M}_{0}$ with the Cattell's threshold set to $0.2$. The
mixing parameter $\alpha$ for kernels was set to $0.5$ in order
not to favor any kernel but it is expected an improvement of the results
if this parameter is tuned too. Kernel parameters have been tuned
by cross-validation on a learning sample and the kernels associated
to these values are presented in Figure~\ref{fig:Mixt-2}. The rows
and columns of the matrices are sorted according to the class memberships
(healthy or sick) and the sick patients are the last ones. We then
compared the performance of pgpDA with the combined kernel to pgpDA
with, on the one hand, a simple RBF kernel built only on the quantitative
variables of the dataset and, on the other hand, a Hamming kernel
built only on the categorical variables. Table~\ref{fig:Mixt-3}
presents both the true positive (TP) and false positive (FP) rates
obtained on 25 replications of the classification experiment for pgpDA
on quantitative data, on categorical data and on the mixed data. It
turns out that quantitative data contains most of the important information
to discriminate the patients with thyroid diseases and that categorical
data, when considered alone, are not enough to build an efficient
classifier. However, it appears that the use of the categorical features
in combination with the quantitative data allows to slightly improve
the prediction of thyroid diseases (increases the TP rate and decreases
the FP rate). In particular, the reduction of the FP rate is important
here since it implies an important reduction of the number of false
alarms.

\begin{table}
\begin{centering}
\begin{tabular}{|c|c|c|c|}
\hline 
Method  & %
\begin{tabular}{c}
pgpDA on\tabularnewline
quantitative data\tabularnewline
\end{tabular} & %
\begin{tabular}{c}
pgpDA on\tabularnewline
categorical data\tabularnewline
\end{tabular} & %
\begin{tabular}{c}
pgpDA with the\tabularnewline
combined kernel\tabularnewline
\end{tabular}\tabularnewline
\hline 
\hline 
TP rate  & 74.86  & 96.00  & 75.88\tabularnewline
\hline 
FP rate  & 22.16  & 95.53  & 21.97\tabularnewline
\hline 
\end{tabular}
\par\end{centering}

\caption{\label{fig:Mixt-3}Classification results on test sets for the Thyroid
dataset (mixed data). Results are averaged on 25 replications of the
experiment.}
\end{table}

\section{Conclusion\label{sec:conclu}}

This work has introduced a family of parsimonious Gaussian process
models for the supervised and unsupervised classification of quantitative
and non-quantitative data. The proposed parsimonious models are obtained
by constraining the eigen-decomposition of the Gaussian processes
modeling each class. They allow in particular to use non-linear mapping
functions which project the observations into an infinite dimensional
space and to build, from a finite sample, a model-based classifier
in this space. It has been also demonstrated that the building of
the classifier can be directly done from the observation space through
a kernel, avoiding the explicit knowledge of the mapping function.
It has been possible to classify data of various nature including
categorical data, functional data, networks and even mixed data by
combining different kernels. The methodology is as well extended to
the unsupervised classification case. Numerical experiments on benchmark
data sets have shown that pgpDA performs similarly or better compared
to the best kernel methods of the state of the art. The possibility
to examine the model parameters and to visualize the data into the
class-specific feature subspaces permits a finer interpretation of
the results than with conventional discriminative kernel methods.
Among the possible extensions of this work, it would be interesting
to extend the methodology to the semi-supervised case in which only
a few observations are labeled.

\appendix

\section*{Appendix: Proofs}

\paragraph{Proof of Proposition~\ref{propun}}

Recalling that $d_{\max}=\max(d_{1},...,d_{k})$, the classification
function can be rewritten as: 
\[
D_{i}(\varphi(x))=\sum_{j=1}^{r}\frac{1}{\lambda_{ij}}<\varphi(x)-\mu_{i},q_{ij}>_{L_{2}}^{2}+\sum_{j=1}^{d_{i}}\log(\lambda_{ij})+\sum_{j=d_{i}+1}^{d_{\max}}\log(\lambda)-2\log(\pi_{i})+\gamma,
\]
 where $\gamma=(r-d_{\max})\log(\lambda)$ is a constant term which
does not depend on the index $i$ of the class. In view of the assumptions,
$D_{i}(\varphi(x))$ can be also rewritten as: 
\begin{eqnarray*}
D_{i}(\varphi(x)) & = & \sum_{j=1}^{d_{i}}\frac{1}{\lambda_{ij}}<\varphi(x)-\mu_{i},q_{ij}>_{L_{2}}^{2}+\frac{1}{\lambda}\sum_{j=d_{i}+1}^{r}<\varphi(x)-\mu_{i},q_{ij}>_{L_{2}}^{2}\\
 & + & \sum_{j=1}^{d_{i}}\log(\lambda_{ij})+(d_{\max}-d_{i})\log(\lambda)-2\log(\pi_{i})+\gamma.
\end{eqnarray*}
Introducing the norm $||.||_{L_{2}}$ associated with the scalar product
$<.,.>_{L_{2}}$ and in view of Proposition~1 of~\cite[p. 208]{Shorack86},
we finally obtain: 
\begin{eqnarray*}
D_{i}(\varphi(x)) & = & \sum_{j=1}^{d_{i}}\left(\frac{1}{\lambda_{ij}}-\frac{1}{\lambda}\right)<\varphi(x)-\mu_{i},q_{ij}>_{L_{2}}^{2}+\frac{1}{\lambda}||\varphi(x)-\mu_{i}||_{L_{2}}^{2}\\
 & + & \sum_{j=1}^{d_{i}}\log(\lambda_{ij})+(d_{\max}-d_{i})\log(\lambda)-2\log(\pi_{i})+\gamma,
\end{eqnarray*}
 which is the desired result.\hfill$\square$

\paragraph{Proof of Proposition~\ref{propdeux}}

The proof involves three steps.

\noindent i) Computation of the projection $<\varphi(x)-\hat{\mu}_{i},\hat{q}_{ij}>_{L_{2}}$
: Since $(\hat{\lambda}_{ij},\hat{q}_{ij})$ is solution of the Fredholm-type
equation, it follows that, for all $t\in[0,1]$, 
\begin{eqnarray}
\hat{\lambda}_{ij}\hat{q}_{ij}(t) & = & \int_{0}^{1}\hat{\Sigma}_{i}(s,t)\hat{q}_{ij}(s)ds\nonumber \\
 & = & \frac{1}{n_{i}}\sum_{x_{\ell}\in C_{i}}<\varphi(x_{\ell})-\hat{\mu}_{i},\hat{q}_{ij}>_{L_{2}}(\varphi(x_{\ell})(t)-\hat{\mu}_{i}(t)).\label{fredholm}
\end{eqnarray}
 This implies that $\hat{q}_{ij}$ lies in the linear subspace spanned
by the $(\varphi(x_{\ell})-\hat{\mu}_{i})$, $x_{\ell}\in C_{i}$.
As a consequence, the rank of the operator $\hat{\Sigma}_{i}$ is
finite and is at most $r_{i}=\min(n_{i},r)$. It therefore exists
$\beta_{ij\ell}\in{\mathbb{R}}$ such that: 
\begin{equation}
\hat{q}_{ij}=\frac{1}{\sqrt{n_{i}\hat{\lambda}_{ij}}}\sum_{x_{\ell}\in C_{i}}\beta_{ij\ell}(\varphi(x_{\ell})-\hat{\mu}_{i})\label{eqvect}
\end{equation}
 leading to: 
\begin{equation}
<\varphi(x)-\hat{\mu}_{i},\hat{q}_{ij}>_{L_{2}}=\frac{1}{\sqrt{n_{i}\hat{\lambda}_{ij}}}\sum_{x_{\ell}\in C_{i}}\beta_{ij\ell}\rho_{i}(x,x_{\ell}),\label{projected-data}
\end{equation}
 for all $j=1,\dots,r_{i}$. The estimated classification function
has therefore the following form: 
\begin{align*}
\hat{D}_{i}(\varphi(x)) & =\frac{1}{n_{i}}\sum_{j=1}^{d_{i}}\frac{1}{\hat{\lambda}_{ij}}\left(\frac{1}{\hat{\lambda}_{ij}}-\frac{1}{\hat{\lambda}}\right)\left(\sum_{x_{\ell}\in C_{i}}\beta_{ij\ell}\rho_{i}(x,x_{\ell})\right)^{2}+\frac{1}{\hat{\lambda}}\rho_{i}(x,x)\\
 & +\sum_{j=1}^{d_{i}}\log(\hat{\lambda}_{ij})+(d_{\max}-d_{i})\log(\hat{\lambda})-2\log(\hat{\pi}_{i}),
\end{align*}
 for all $i=1,\dots,k$.

\noindent ii) Computation of the $\beta_{ij\ell}$ and $\hat{\lambda}_{ij}$:
Replacing (\ref{eqvect}) in the Fredholm-type equation (\ref{fredholm})
it follows that 
\begin{eqnarray*}
\frac{1}{n_{i}}\sum_{x_{\ell},x_{\ell'}\in C_{i}}\beta_{ij\ell'}(\varphi(x_{\ell})-\hat{\mu}_{i})\rho_{i}(x_{\ell},x_{\ell'})=\hat{\lambda}_{ij}\sum_{x_{\ell'}\in C_{i}}\beta_{ij\ell'}(\varphi(x_{\ell'})-\hat{\mu}_{i}).
\end{eqnarray*}
 Finally, projecting this equation on $\varphi(x_{m})-\hat{\mu}_{i}$
for $x_{m}\in C_{i}$ yields 
\begin{eqnarray*}
\frac{1}{n_{i}}\sum_{x_{\ell},x_{\ell'}\in C_{i}}\beta_{ij\ell'}\rho_{i}(x_{\ell},x_{m})\rho_{i}(x_{\ell},x_{\ell'})=\hat{\lambda}_{ij}\sum_{x_{\ell'}\in C_{i}}\beta_{ij\ell'}\rho_{i}(x_{m},x_{\ell'}).
\end{eqnarray*}
 Recalling that $M_{i}$ is the matrix $n_{i}\times n_{i}$ defined
by $(M_{i})_{\ell,\ell'}=\rho_{i}(x_{\ell},x_{\ell'})/n_{i}$ and
introducing $\beta_{ij}$ the vector of ${\mathbb{R}}^{n_{i}}$ defined
by $(\beta_{ij})_{\ell}=\beta_{ij\ell}$, the above equation can be
rewritten as $M_{i}^{2}\beta_{ij}=\hat{\lambda}_{ij}M_{i}\beta_{ij}$
or, after simplification $M_{i}\beta_{ij}=\hat{\lambda}_{ij}\beta_{ij}.$
As a consequence, $\hat{\lambda}_{ij}$ is the $j$th largest eigenvalue
of $M_{i}$ and $\beta_{ij}$ is the associated eigenvector for all
$1\leq j\leq d_{i}$. Let us note that the constraint $\|\hat{q}_{ij}\|=1$
can be rewritten as $\beta_{ij}^{t}\beta_{ij}=1$.

\noindent iii) Computation of $\hat{\lambda}$: Remarking that trace$(\hat{\Sigma}_{i})=\mbox{trace\ensuremath{(M_{i})+\mbox{\ensuremath{\sum_{j=r_{i}+1}^{r}\hat{\lambda}_{ij}}}}}$,
it follows: 
\[
\hat{\lambda}=\frac{1}{\sum_{i=1}^{k}\hat{\pi_{i}}(r_{i}-d_{i})}\sum_{i=1}^{k}\hat{\pi}_{i}\left(\mathrm{trace}(M_{i})-\sum_{j=1}^{d_{i}}\hat{\lambda}_{ij}\right),
\]
 and the proposition is proved.\hfill$\square$

\paragraph{Proof of Proposition~\ref{proptrois}}

It is sufficient to prove that $\hat{q}_{ij}$ and $\hat{\lambda}_{ij}$
are respectively the $j$th normed eigenvector and eigenvalue of $\hat{\Sigma}_{i}$.
First, 
\begin{align*}
\hat{\Sigma}_{i}\hat{q}_{ij} & =\frac{1}{\sqrt{n_{i}\hat{\lambda}_{ij}}}\frac{1}{n_{i}}\sum_{x_{\ell'}\in C_{i}}(x_{\ell'}-\bar{\mu}_{i})(x_{\ell'}-\bar{\mu}_{i})^{t}\sum_{x_{\ell}\in C_{i}}\beta_{ij\ell}(x_{\ell}-\bar{\mu}_{i})\\
 & =\frac{1}{\sqrt{n_{i}\hat{\lambda}_{ij}}}\frac{1}{n_{i}}\sum_{x_{\ell'},x_{\ell}\in C_{i}}\beta_{ij\ell}(x_{\ell'}-\bar{\mu}_{i})(x_{\ell'}-\bar{\mu}_{i})^{t}(x_{\ell}-\bar{\mu}_{i})\\
 & =\frac{1}{\sqrt{n_{i}\hat{\lambda}_{ij}}}\frac{1}{n_{i}}\sum_{x_{\ell'},x_{\ell}\in C_{i}}\beta_{ij\ell}(x_{\ell'}-\bar{\mu}_{i})\rho_{i}(x_{\ell},x_{\ell'})\\
 & =\frac{1}{\sqrt{n_{i}\hat{\lambda}_{ij}}}\sum_{x_{\ell'},x_{\ell}\in C_{i}}(M_{i})_{\ell,\ell'}\beta_{ij\ell}(x_{\ell'}-\bar{\mu}_{i})\\
 & =\frac{1}{\sqrt{n_{i}\hat{\lambda}_{ij}}}B^{-1}\sum_{x_{\ell'}\in C_{i}}(M_{i}\beta_{ij})_{\ell'}(x_{\ell'}-\bar{\mu}_{i}),
\end{align*}
 and remarking that $\beta_{ij}$ is eigenvector of $M_{i}$, it follows:
\[
\hat{\Sigma}_{i}\hat{q}_{ij}=\hat{\lambda}_{ij}\frac{1}{\sqrt{n_{i}\hat{\lambda}_{ij}}}B^{-1}\sum_{x_{\ell'}\in C_{i}}\beta_{ij\ell'}(x_{\ell'}-\bar{\mu}_{i})=\hat{\lambda}_{ij}\hat{q}_{ij}.
\]
 Second, straightforward algebra shows that 
\begin{align*}
||\hat{q}_{ij}||^{2} & =\frac{1}{n_{i}\hat{\lambda}_{ij}}\sum_{x_{\ell}\in C_{i}}\beta_{ij\ell}(x_{\ell})-\bar{\mu}_{i})^{t}\sum_{x_{\ell'}\in C_{i}}\beta_{ij\ell'}(x_{\ell'}-\bar{\mu}_{i})\\
 & =\frac{1}{n_{i}\hat{\lambda}_{ij}}\sum_{x_{\ell'},x_{\ell}\in C_{i}}\beta_{ij\ell}\beta_{ij\ell'}(x_{\ell}-\bar{\mu}_{i})^{t}(x_{\ell'}-\bar{\mu}_{i})\\
 & =\frac{1}{\hat{\lambda}_{ij}}\sum_{x_{\ell'},x_{\ell}\in C_{i}}(M_{i})_{\ell,\ell'}\beta_{ij\ell}\beta_{ij\ell'}\\
 & =\frac{1}{\hat{\lambda}_{ij}}\hat{q}_{ij}^{t}M_{i}\hat{q}_{ij}=1,
\end{align*}
 and the result is proved.\hfill$\square$

\paragraph{Proof of Proposition~\ref{propquatre}}

For all $\ell=1,\dots,L$, the $\ell$th coordinate of the mapping
function $\varphi(x)$ is defined as the $\ell$th coordinate of the
function $x$ expressed in the truncated basis $\{b_{1},\dots,b_{L}\}$.
More specifically, 
\[
x(t)=\sum_{\ell=1}^{L}\varphi_{\ell}(x)b_{\ell}(t),
\]
 for all $t\in[0,1]$ and thus, for all $j=1,\dots,L$, we have 
\[
\gamma_{j}(x)=\int_{0}^{1}x(t)b_{j}(t)dt=\sum_{\ell=1}^{L}\varphi_{\ell}(x)\int_{0}^{1}b_{j}(t)b_{\ell}(t)dt=\sum_{\ell=1}^{L}B_{j\ell}\varphi_{\ell}(x).
\]
 As a consequence, $\varphi(x)=B^{-1}\gamma(x)$ and $K(x,y)=\gamma(x)^{t}B^{-1}\gamma(y)$.
Introducing 
\[
\bar{\gamma}_{i}=\frac{1}{n_{i}}\sum_{x_{j}\in C_{i}}\gamma(x_{j}),
\]
 it follows that $\rho_{i}(x,y)=(\gamma(x)-\bar{\gamma}_{i})^{t}B^{-1}(\gamma(y)-\bar{\gamma}_{i})$.
Let us first show that $\hat{q}_{ij}$ is eigenvector of $B^{-1}\hat{\Sigma}_{i}$.
Recalling that 
\[
\hat{q}_{ij}=\frac{1}{\sqrt{n_{i}\hat{\lambda}_{ij}}}B^{-1}\sum_{x_{\ell}\in C_{i}}\beta_{ij\ell}(\gamma(x_{\ell})-\bar{\gamma}_{i}),
\]
 we have 
\begin{align*}
B^{-1}\hat{\Sigma}_{i}\hat{q}_{ij} & =\frac{1}{\sqrt{n_{i}\hat{\lambda}_{ij}}}B^{-1}\frac{1}{n_{i}}\sum_{x_{\ell'}\in C_{i}}(\gamma(x_{\ell'})-\bar{\gamma}_{i})(\gamma(x_{\ell'})-\bar{\gamma}_{i})^{t}B^{-1}\sum_{x_{\ell}\in C_{i}}\beta_{ij\ell}(\gamma(x_{\ell})-\bar{\gamma}_{i})\\
 & =\frac{1}{\sqrt{n_{i}\hat{\lambda}_{ij}}}B^{-1}\frac{1}{n_{i}}\sum_{x_{\ell'},x_{\ell}\in C_{i}}\beta_{ij\ell}(\gamma(x_{\ell'})-\bar{\gamma}_{i})(\gamma(x_{\ell'})-\bar{\gamma}_{i})^{t}B^{-1}(\gamma(x_{\ell})-\bar{\gamma}_{i})\\
 & =\frac{1}{\sqrt{n_{i}\hat{\lambda}_{ij}}}B^{-1}\frac{1}{n_{i}}\sum_{x_{\ell'},x_{\ell}\in C_{i}}\beta_{ij\ell}(\gamma(x_{\ell'})-\bar{\gamma}_{i})\rho_{i}(x_{\ell},x_{\ell'})\\
 & =\frac{1}{\sqrt{n_{i}\hat{\lambda}_{ij}}}B^{-1}\sum_{x_{\ell'},x_{\ell}\in C_{i}}(M_{i})_{\ell,\ell'}\beta_{ij\ell}(\gamma(x_{\ell'})-\bar{\gamma}_{i})\\
 & =\frac{1}{\sqrt{n_{i}\hat{\lambda}_{ij}}}B^{-1}\sum_{x_{\ell'}\in C_{i}}(M_{i}\beta_{ij})_{\ell'}(\gamma(x_{\ell'})-\bar{\gamma}_{i}).
\end{align*}
 Remarking that $\beta_{ij}$ is eigenvector of $M_{i}$, it follows:
\[
B^{-1}\hat{\Sigma}_{i}\hat{q}_{ij}=\hat{\lambda}_{ij}\frac{1}{\sqrt{n_{i}\hat{\lambda}_{ij}}}B^{-1}\sum_{x_{\ell'}\in C_{i}}\beta_{ij\ell'}(\gamma(x_{\ell'})-\bar{\gamma}_{i})=\hat{\lambda}_{ij}\hat{q}_{ij}.
\]
 Let us finally compute the norm of $\hat{q}_{ij}$: 
\begin{align*}
||\hat{q}_{ij}||^{2} & =\frac{1}{n_{i}\hat{\lambda}_{ij}}\sum_{x_{\ell}\in C_{i}}\beta_{ij\ell}(\gamma(x_{\ell})-\bar{\gamma}_{i})^{t}B^{-1}\sum_{x_{\ell'}\in C_{i}}\beta_{ij\ell'}(\gamma(x_{\ell'})-\bar{\gamma}_{i})\\
 & =\frac{1}{n_{i}\hat{\lambda}_{ij}}\sum_{x_{\ell'},x_{\ell}\in C_{i}}\beta_{ij\ell}\beta_{ij\ell'}(\gamma(x_{\ell})-\bar{\gamma}_{i})^{t}B^{-1}(\gamma(x_{\ell'})-\bar{\gamma}_{i})\\
 & =\frac{1}{\hat{\lambda}_{ij}}\sum_{x_{\ell'},x_{\ell}\in C_{i}}(M_{i})_{\ell,\ell'}\beta_{ij\ell}\beta_{ij\ell'}\\
 & =\frac{1}{\hat{\lambda}_{ij}}\hat{q}_{ij}^{t}M_{i}\hat{q}_{ij}=1,
\end{align*}
 and the result is proved.\hfill$\square$

\bibliographystyle{plain}
\bibliography{bib}

\end{document}